\documentclass[%
 reprint,
 amsmath,amssymb,
 aps,
]{revtex4-2}

\usepackage{graphicx}
\usepackage{dcolumn}
\usepackage{bm}

\usepackage{color}     

\begin{document}

\preprint{APS/123-QED}

\title{Description of $^{93}$Nb stellar electron-capture rates by the Projected Shell Model}

\author{Long-Jun Wang}
\email{longjun@swu.edu.cn}
\altaffiliation{School of Physical Science and Technology, Southwest University, Chongqing 400715, China} 
\author{Liang Tan}%
\affiliation{School of Physical Science and Technology, Southwest University, Chongqing 400715, China}%
\author{Zhipan Li}%
\affiliation{School of Physical Science and Technology, Southwest University, Chongqing 400715, China}%

\author{Bingshui Gao}%
\affiliation{CAS Key Laboratory of High Precision Nuclear Spectroscopy, Institute of Modern Physics, Chinese Academy of Sciences, Lanzhou 730000, China }
\affiliation{School of Nuclear Science and Technology, University of Chinese Academy of Sciences, Beijing 100049, China}

\author{Yang Sun}
\affiliation{School of Physics and Astronomy, Shanghai Jiao Tong University, Shanghai 200240, China}%

\date{\today}

\begin{abstract}
  Capture of electrons by nuclei is an important process in stellar environments where excited nuclear states are thermally populated. However, accurate treatment for excited configurations in electron capture (EC) rates has been an unsolved problem for medium-heavy and heavy nuclei. In this work, we take the $^{93}$Nb $\rightarrow$ $^{93}$Zr EC rates as the example to introduce the Projected-Shell-Model (PSM) in which excited configurations are explicitly included as multi-quasiparticle states. Applying the prevalent assumption that the parent nucleus always stays in its ground state in stellar conditions, we critically compare the obtained PSM results with the recently-measured Gamow-Teller transition data, and with the previous calculations by the conventional shell model and the quasiparticle random-phase approximation. We discuss important ingredients that are required in theoretical models used for stellar EC calculations, and demonstrate effects of the explicit inclusion of excited nuclear states in EC rate calculations, especially when both electron density and environment temperature are high.
\end{abstract}

\maketitle


\section{\label{sec:intro}introduction}

The understanding of the dynamics of some astrophysical
objects relies on reliable information on the electron capture (EC) process  \cite{Fuller1, Fuller2, Fuller3, Fuller4, Heger_2001_PRL, Hix_2003_PRL, Langanke-RMP, langanke_2020_EC_review}, 
\begin{eqnarray} \label{eq:EC_general}
  ^{A}_{Z} \mathrm{X} + e^- \rightarrow \  ^{\ \  A}_{Z-1} \mathrm{Y} +\nu_e,
\end{eqnarray}
in which the parent nucleus (with proton number $Z$ and mass number $A$) captures a relativistic electron and decays to the daughter nucleus by emitting a neutrino. Due to the nature of the stellar EC reaction, it may lead to three important consequences with regard to compact astrophysical objects: reducing the degeneracy pressure of electrons, cooling the environment by neutrino escaping, and driving nuclei to be more neutron-rich (see Ref. \cite{langanke_2020_EC_review} for the latest review). Especially, the information of EC process is considered to be crucial for core-collapse supernovae \cite{Langanke-RMP, physrep-2007, cole-2012, EC_93Nb_PRC2020}, cooling of neutron star crust and ocean \cite{Nature2014, Our_Urca_arXiv}, and the thermonuclear explosions of accreting white dwarfs \cite{Iwamoto_1999}, etc.

Electron capture is one of the fundamental nuclear processes
mediated by the weak interaction. Stellar EC rates can be theoretically determined following the pioneering works by Fuller, Fowler, and Newman (FFN) \cite{Fuller1, Fuller2, Fuller3, Fuller4}, where allowed nuclear transitions (especially the Gamow-Teller transitions in the $\beta^+$ direction) dominate. In principle, the Gamow-Teller (GT) transition-strength distributions can be measured with the modern experimental technique with the  charge-exchange (CE) reactions \cite{Zegers2008prc, Fujita-2011-ppnp, EC_93Nb_PRC2020, BSGao_2021_PRL}, although the energy resolution is relatively limited at present. One should however keep in mind that stellar electron captures may differ
substantially from those which can be studied in laboratory today. For many relevant astrophysical simulations such as the core-collapse supernovae and neutron star crust, stellar EC rates for a large number of nuclei are required. Furthermore, the contributions from excited states of parent nuclei are indispensable \cite{Heger_2001_PRL, Tan_2020_PLB}. Therefore, for majority of the cases one has to rely on theoretical calculations. 

In the modern treatment, several nuclear-structure models have been applied to study the GT transition and stellar EC reactions. For the $sd$-shell (with $A=17-39$) \cite{Oda-1994, Gabriel-2014, wendell-2014} and some $pf$-shell ($A=45-65$) nuclei \cite{Caurier-1999, Langanke2000, Langanke2001, cole-2012}, the large-scale shell-model (LSSM) diagonalization method is considered as the optimal tool \cite{Langanke-RMP, langanke_2020_EC_review}. While this method is exact within the adopted model space, it becomes physically inappropriate when environment temperature is high. This is because when highly excited nuclear states are involved in the calculation, cross-shell configurations are 
inevitably needed, which is technically difficult for the LSSM. For medium-heavy and heavy nuclei ($A>65$) especially if they are deformed, alternative methods 
must be applied. The popular methods include the hybrid model based on the shell model Monte Carlo approach and the random-phase approximation \cite{Langanke_2001_PRC, Langanke_2003_PRL}, and the quasiparticle random-phase approximation (QRPA) with different energy density functionals \cite{jon-1999, ec-Paar, Bai-2010, Dzhioev2010, ec-Niu, fang-2013, Niu-2013, Sarriguren2013, Robin-2019, Suhonen-2019-review}. These methods differ from each other in model space (the size of single-particles in the intrinsic mean field) and/or in truncation on configuration space (the variety of many-body correlations among valence nucleons beyond the mean field), as well as in effective (two-body) interactions. While these methods usually do not have problem for including a large model space, the way to build their configuration space does matter for stellar EC calculations. This is because even that the sumrule for the total GT transitions is satisfied, the GT fragmentation as a function of excited states, which are closely related to  model configurations, can sensitively influence the stellar EC results.

The stellar EC rates for medium-heavy, odd-mass nuclei exhibit some unique characteristics as compared to those in even-number nuclei. The main challenge in odd-mass calculations lies in the fact of dense and variety in the  distribution of single-particle-type of states near the ground state. Furthermore, unlike in even-number nuclei where the $0^+ \rightarrow 1^+$ GT transition is the main concern, in odd-mass nuclei, spin and parity for individual low-lying levels determine the GT distributions due to the selection rule of allowed GT transition. For example, for the $^{93}$Nb $\rightarrow$ $^{93}$Zr case, the ground state of the parent nucleus has the spin-parity $9/2^+$ (see Fig. \ref{fig:one}) and the GT transitions from such a ground state would populate many final states with spin-parities of $\{7/2^+, 9/2^+, 11/2^+ \}$ of the daughter nucleus. Transitions from individual excited states of the parent to the distributed states of the daughter can result in a rich variety of GT patterns, for which the Brink-Axel hypothesis \cite{Brink1955, Axel1962} may hardly be expected as a good approximation.

In the present work, we provide the calculation by the Projected Shell Model for stellar EC rates of medium-heavy odd-mass nuclei, taking $^{93}$Nb $\rightarrow$ $^{93}$Zr as the first example. When only transitions from the ground state of the parent nucleus are considered, our PSM results are tested by the very recent data and are compared to the results of the LSSM and the QRPA. For more realistic stellar EC rates where transitions from all excited states of the parent nucleus are taken into account, the contribution and effect of excited states are studied, and discussed for different densities and temperatures of the stellar environment. In Sec. \ref{sec:theory} we briefly introduce our PSM method for the calculation of stellar EC rates. In Sec. \ref{sec:result} we compare our calculations for the EC reaction of $^{93}$Nb $\rightarrow$ $^{93}$Zr with the recent data and with previous results of other nuclear models. We finally summarize our work in Sec. \ref{sec:sum}. 

\section{\label{sec:theory}Theoretical framework}

For high-density and high-temperature stellar environments, we assume that parent nuclei are in a thermal equilibrium with occupation probability for excited states following the Boltzmann distribution. The stellar EC rates (in s$^{-1}$) can be expressed as \cite{Fuller1, Fuller2, Fuller3, Fuller4}
\begin{eqnarray} \label{eq.lambda-EC}
  \lambda^{\text{EC}} = \frac{\ln 2}{K} \sum_i \frac{(2J_i+1)e^{-E_i/(k_BT)}} {G(Z, A, T)} 
  \sum_f B_{if} \Phi^{\text{EC}}_{if} ,
\end{eqnarray}
which explicitly takes into account all relevant transitions from initial ($i$) states of EC parent nucleus to final ($f$) states of  daughter nucleus. In Eq. (\ref{eq.lambda-EC}), the first summation over $i$ represents the occupation probability for the parent-nuclear states at specific stellar temperature $T$, where $G(Z, A, T) = \sum_i (2J_i+1) \exp (-E_i/(k_B T))$ is the partition function with $k_B$ being the Boltzmann constant. $J_i$ and $E_i$ appearing in the $i$-summation in Eq. (\ref{eq.lambda-EC}) are respectively the angular momenta and excitation energies of the parent nucleus. The constant $K$ in Eq. (\ref{eq.lambda-EC}) can be determined from superallowed Fermi transitions, and $K=6146 \pm 6$ s \cite{Haxton-book} is adopted here.  $\Phi^{\text{EC}}_{if}$ in the second summation over $f$ is the phase space integral for individual nuclear transitions (with corresponding reduced transition probability $B_{if}$),  
\begin{eqnarray} \label{eq.Phi}
  \Phi^{\text{EC}}_{if} = \int_{\omega_l}^\infty \omega p (Q_{if}+\omega)^2 F(Z, \omega) S_e(\omega) d\omega,
\end{eqnarray}
where $\omega$ and $p=\sqrt{\omega^2-1}$ (in $m_e c^2$ and $m_ec$, respectively) label the total (rest mass and kinetic) energy and momentum of the electron. For individual transitions, the total available energy (the $Q_{if}$-value) is 
\begin{eqnarray}
  Q_{if} = \frac{1}{m_ec^2} (M_p - M_d + E_i - E_f),
\end{eqnarray}
where $M_p$ ($M_d$) denotes the nuclear mass of the parent (daughter) nucleus and $E_f$ is the excitation energy of the daughter nucleus. $\omega_l$ labels the capture threshold energy (in $m_e c^2$), with $\omega_l = 1$ if $Q_{if} > -1$, or $\omega_l = |Q_{if}|$ if $Q_{if} < -1$. For a given stellar temperature, the distribution functions read as
\begin{eqnarray} \label{eq.se}
  S_{e/p}(\omega) = \frac{1}{\exp{[(\omega \mp \mu_e) / k_B T)] + 1}} ,
\end{eqnarray}
for electron (``$-$") and positron (``$+$"), where the electron chemical potential $\mu_e$ could be determined with the provided stellar baryon density $\rho$, the electron-to-baryon ratio $Y_e$ through
\begin{eqnarray} \label{eq.rhoye}
  \rho Y_e = \frac{1}{\pi^2 N_A} \Big(\frac{m_e c}{\hbar} \Big)^3 \int _0 ^\infty (S_e - S_p) p^2 dp,
\end{eqnarray}
with $N_A$ the Avogadro's number. In Eq. (\ref{eq.Phi}), $F(Z, \omega)$ is the Fermi function that reflects the Coulomb distortion of the electron wave function near the nucleus \cite{Fuller1, Langanke2000}.

In calculating stellar EC rates, the central task is the evaluation of $B_{if}$ in the second summation over $j$ in Eq. (\ref{eq.lambda-EC}). In the present work we consider the stellar environments with the typical densities $\rho Y_e = 10^7, 10^9$, and $10^{11}$ g/cm$^3$, and temperatures $T=1 - 15$ GK ($10^9$ K). Under these conditions, the allowed Gamow-Teller transitions dominate. The reduced transition probability, $B(\text{GT})$, being defined such that the strength associated with the decay of the free neutron has $B(\text{GT}) = 3$, can be measured experimentally with the modern CE reactions \cite{EC_93Nb_PRC2020}, or calculated by nuclear-structure models as 
\begin{eqnarray} \label{eq.BGT}
  B_{if}(\text{GT}^+) = \Big(\frac{g_A}{g_V}\Big)^2_{\text{eff}} \frac{ \big\langle \Psi^{n_f}_{J_f} \big\| \sum_k \hat\sigma^k \hat\tau^k_+ \big\| \Psi^{n_i}_{J_i} \big\rangle ^2}{2J_i+1}
\end{eqnarray}
where $(g_A/g_V)_{\text{eff}}$ is the effective ratio of axial and vector coupling constants,  
\begin{eqnarray} \label{eq.quench}
  \Big(\frac{g_A}{g_V}\Big)_{\text{eff}} = f_{\text{quench}} \Big(\frac{g_A}{g_V}\Big)_{\text{bare}} ,
\end{eqnarray}
which usually differs from the bare value  $(g_A/g_V)_{\text{bare}}=-1.2599(25)$ \cite{Haxton-book} or $= -1.27641(45)$ \cite{new_gAgV} by a quenching factor. In the present $^{93}$Nb $\rightarrow$ $^{93}$Zr work we take $f_{\text{quench}} = 0.9$ as suggested in Refs. \cite{Javier2011PRL, Wang-0vbb-2018, Nature2019, quenching_2020_PLB} for medium-heavy nuclei. The GT transition operator in Eq. (\ref{eq.BGT}) keeps only the one-body current contributions as in the chiral effective field theory \cite{ChiralEFT2009, Javier2011PRL, Klos_2017_EPJA, Wang-0vbb-2018}, which includes the Pauli spin operator $\hat\sigma$ and the isospin raising operator $\hat\tau_+$ for electron captures. 

Description of the nuclear many-body states, $\Psi^{n}_{J}$ in Eq. (\ref{eq.BGT}), is purely a structure problem, and is precisely the place where nuclear structure enters into the discussion. It is desired that  calculations for excited states in  medium-heavy nuclei apply modern many-body techniques to obtain properly the wave functions with good angular momentum $J$, as this index appears explicitly in Eq. (\ref{eq.BGT}). In the present work such wave functions are described in the PSM by the projection technique \cite{PSM-Sun, Gao2006, Wang-2014-R, Wang_2016_PRC, PSM-Sun2, Wang2018}, i.e., 
\begin{eqnarray}\label{Eq.wave-func}
  |\Psi^n_{JM}\rangle = \sum_{K\kappa} F^n_{J K\kappa} \hat P^J_{MK} |\Phi_\kappa\rangle ,
\end{eqnarray}
where the angular-momentum projection operator $\hat P^J_{MK}$ (see Refs. \cite{many-body, PSM-review} for details) restores the broken rotational symmetry in the intrinsic frame and transforms the description of the wave functions from the intrinsic to the laboratory frame. In the present work, the employed multi-quasiparticle (qp) configuration space, $|\Phi_\kappa\rangle$ in Eq. (\ref{Eq.wave-func}), is large enough, which includes up to 7-qp states, i.e.
\begin{align}\label{Eq.config-n}
  \Big\{ & \hat a^\dag_{\nu_i}|\Phi\rangle, \ \hat a^\dag_{\nu_i} \hat a^\dag_{\nu_j} \hat a^\dag_{\nu_k}|\Phi \rangle,
           \ \hat a^\dag_{\nu_i} \hat a^\dag_{\pi_j} \hat a^\dag_{\pi_k}|\Phi\rangle , \nonumber \\
  & \hat a^\dag_{\nu_i} \hat a^\dag_{\nu_j} \hat a^\dag_{\nu_k} \hat a^\dag_{\pi_l} \hat a^\dag_{\pi_m} |\Phi \rangle,
    \ \hat a^\dag_{\nu_i} \hat a^\dag_{\nu_j} \hat a^\dag_{\nu_k} \hat a^\dag_{\nu_l} \hat a^\dag_{\nu_m} \hat a^\dag_{\pi_n}
      \hat a^\dag_{\pi_o} |\Phi \rangle, \nonumber \\
  & \hat a^\dag_{\nu_i} \hat a^\dag_{\nu_j} \hat a^\dag_{\nu_k} \hat a^\dag_{\pi_l} \hat a^\dag_{\pi_m} \hat a^\dag_{\pi_n}
    \hat a^\dag_{\pi_o} |\Phi \rangle.  \Big\}
\end{align}
for odd-neutron nuclei and 
\begin{align}\label{Eq.config-p}
  \Big\{ & \hat a^\dag_{\pi_i}|\Phi\rangle , \ \hat a^\dag_{\pi_i} \hat a^\dag_{\pi_j} \hat a^\dag_{\pi_k}|\Phi \rangle,
         \ \hat a^\dag_{\pi_i} \hat a^\dag_{\nu_j} \hat a^\dag_{\nu_k}|\Phi \rangle, \nonumber \\
         & \hat a^\dag_{\pi_i} \hat a^\dag_{\pi_j} \hat a^\dag_{\pi_k} \hat a^\dag_{\nu_l} \hat a^\dag_{\nu_m} |\Phi \rangle, \
           \hat a^\dag_{\pi_i} \hat a^\dag_{\pi_j} \hat a^\dag_{\pi_k} \hat a^\dag_{\pi_l} \hat a^\dag_{\pi_m} \hat a^\dag_{\nu_n}
           \hat a^\dag_{\nu_o} |\Phi \rangle, \nonumber \\
         & \hat a^\dag_{\pi_i} \hat a^\dag_{\pi_j} \hat a^\dag_{\pi_k} \hat a^\dag_{\nu_l} \hat a^\dag_{\nu_m} \hat a^\dag_{\nu_n}
           \hat a^\dag_{\nu_o} |\Phi \rangle.  \Big\}
\end{align}
for odd-proton nuclei. In Eqs. (\ref{Eq.config-n}) and (\ref{Eq.config-p}), $|\Phi\rangle$ is the qp vacuum state, and $\hat a^\dag_\nu$ ($\hat a^\dag_\pi$) the  corresponding neutron (proton) qp creation operators, which are obtained by the Nilsson+BCS mean-field calculations. For the detailed construction of the wave function and the philosophy of the model, we refer the recent review article \cite{PSM-Sun2}.

The Hamiltonian is diagonalized by solving the Hill-Wheeler-Griffin equation for the case of a  non-orthonormal basis as in the PSM, which determines the expansion coefficients $F^n_{J K\kappa}$ in Eq. (\ref{Eq.wave-func}). In this work, the separable two-body Hamiltonian with explicit two-body GT force as well as the same numerical parameter as in Refs. \cite{Wang2018, Tan_2020_PLB} are adopted. The Nilsson parameters are taken from Ref. \cite{Nilsson_RMP} and the quadrupole deformation is adopted as $\varepsilon_2 = 0.1$ for both nuclei. 

The present PSM calculation for $A=93$ nuclei considers a large model space, allowing the nucleons active in four major harmonic-oscillator shells ($N=2,3,4,5$ with $^{16}$O as the inert core). Such a large model space ensures that both the neutron and proton Fermi levels lie roughly in the middle of the single-particle space, so that the nucleons have sufficient freedom to be excited. However, the unique feature for the PSM, which differ conceptually from all other existing structure models for stellar EC rates, is the algorithm for the construction of the configuration space. As shown in Eqs. (\ref{Eq.config-n}) and (\ref{Eq.config-p}), we classify the configuration space by the $(2n+1)$-qp states (with $n$ being integers), corresponding to $n$ broken-pairs in each configuration. For example, the term $\hat a^\dag_{\nu_i} \hat a^\dag_{\pi_j} \hat a^\dag_{\pi_k}|\Phi \rangle$ in Eq. (\ref{Eq.config-n}) describes a 3-qp configuration in an odd-neutron nucleus with one broken proton-pair added to the single-neutron state. This algorithm for constructing the PSM configuration space using the broken-pair states as building blocks was proposed to study nuclear chaoticity at highly excited regions \cite{Wang_2019_JpG, Wang_2020_PLB_chaos}, and is supported by the Oslo experiments for nuclear level density, where, as excitation energies go up, clear evidences for step-wise pair breakings are found \cite{Guttormsen_2015_EPJA}. The observed several peaks in the $^{93}$Nb $\rightarrow$ $^{93}$Zr GT distribution \cite{EC_93Nb_PRC2020} seem to indicate also the pair breaking mechanism in the daughter nucleus although the current experimental resolution may not allow a quantitative interpretation.

\section{\label{sec:result}results and discussions}

With the method introduced in the previous section, we calculate stellar EC rates taking the $^{93}$Nb $\rightarrow$ $^{93}$Zr case as the example to discuss features of the PSM results. The discussion begins with the level structure of the parent and daughter nuclei and the characteristic GT strength distribution, before the EC rates at different densities and temperatures are presented. The theoretical results from the PSM are illustrated in parallel with those from two other models, LSSM and QRPA, and compared with the recent experimental data in Ref. \cite{EC_93Nb_PRC2020} measured at MSU. 

\subsection{Level structure in parent and daughter nuclei}

Level structures in parent and daughter nuclei directly influence calculated EC rates as their labels of the nuclear states appear explicitly in the two summations for $\lambda^{\rm EC}$ in Eq. (\ref{eq.lambda-EC}). Especially, the level distribution determines patterns of GT strengths because a GT strength distribution is expressed as the function of excitation energies of the daughter nucleus. Therefore, for a model that is used for calculation of EC rates, it is required that it can correctly describe basic features of the level structure.   

\begin{figure}[htbp]
\begin{center}
  \includegraphics[width=0.48\textwidth]{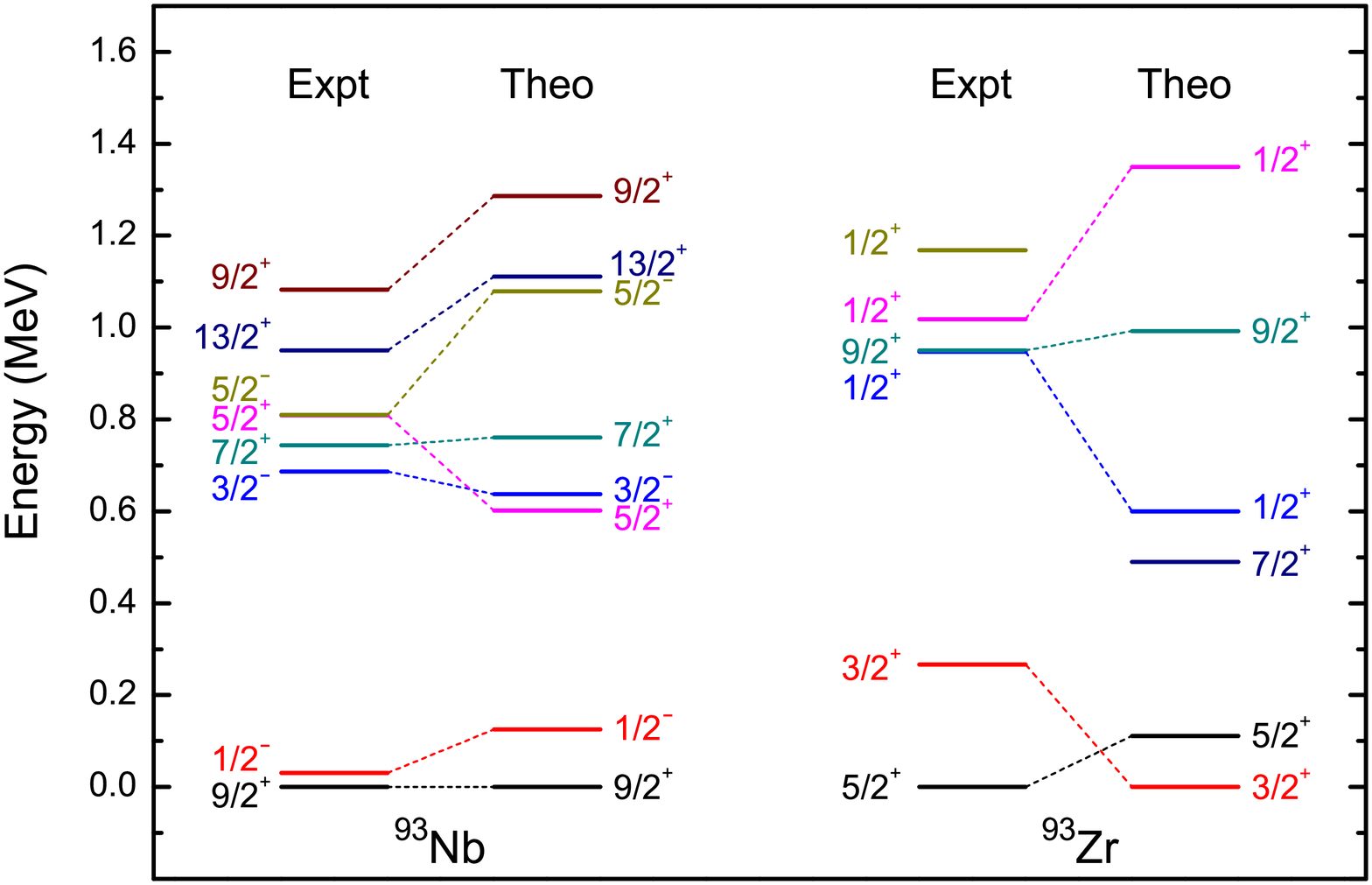}
  \caption{\label{fig:one} (Color online) The calculated energy levels for $^{93}$Nb and $^{93}$Zr, and compared with experimental data \cite{93_level_data}.}
\end{center}
\end{figure}

\begin{figure*}[htbp]
\begin{center}
  \includegraphics[width=0.80\textwidth]{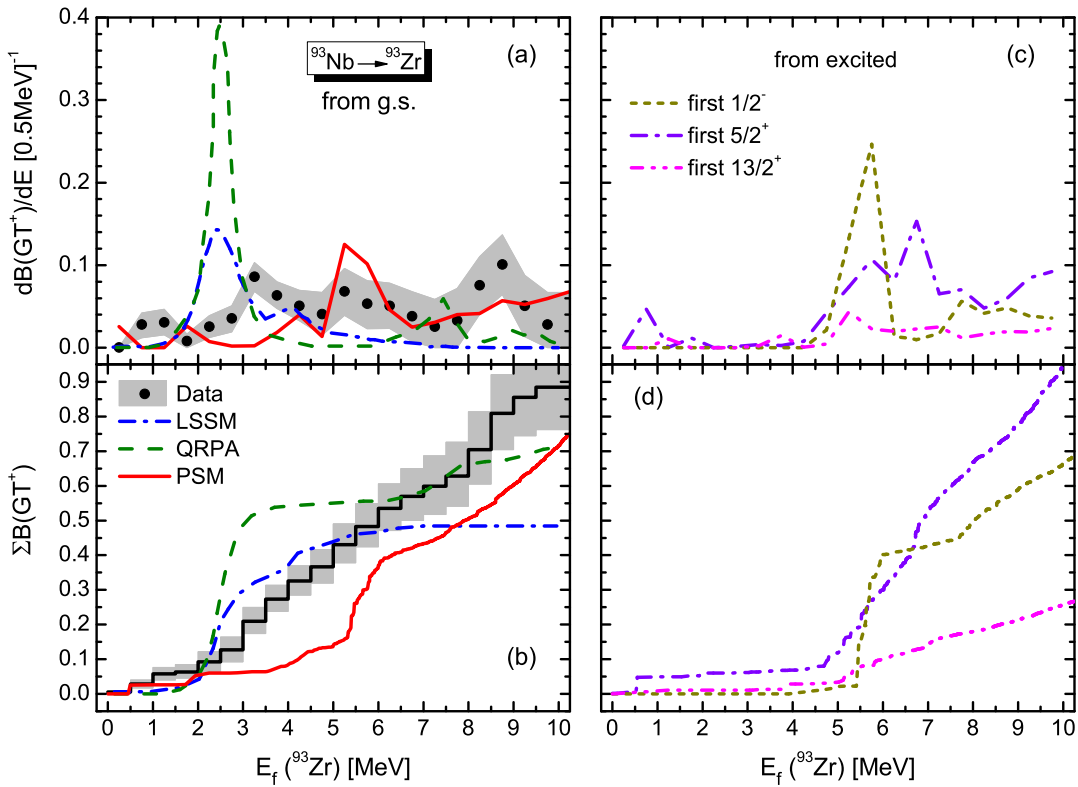}
  \caption{\label{fig:two} (Color online) The GT strength distribution $B(\mathrm{GT}^+)$ (upper panels) and the cumulative sum of the $B(\mathrm{GT}^+)$ (lower panels) for the transitions from $^{93}$Nb to $^{93}$Zr as a function of the excitation energy of the daughter nucleus $^{93}$Zr. For transitions from the ground state (g.s.) of $^{93}$Nb (panels (a) and (b)), calculations by three nuclear models are compared with each other and the recent data \cite{EC_93Nb_PRC2020}. For transitions from different excited states of $^{93}$Nb, predictions of the projected shell model are shown in panels (c) and (d). See the text for details. }
\end{center}
\end{figure*}

In Fig. \ref{fig:one}, we show the PSM calculation for the EC parent ($^{93}$Nb, with $Z=41$ and $N=52$) and the daughter nucleus ($^{93}$Zr, with $Z=40$ and $N=53$) for energy levels up to $E_{i/f} \lesssim 1.2$ MeV, and compare them with the known experimental data \cite{93_level_data}. It is seen that for $^{93}$Nb, the energy levels are described reasonably well by the calculation. The spin-parity of the ground-state is correctly reproduced as $9/2^+$, with the corresponding wave function found to be a strong mixture of proton 1-qp states originating from the proton $g_{9/2}$ orbital. The very low-lying first excited negative-parity $1/2^-$ state is also well described. Its wave function is found to be rather pure, with the main configuration as $\pi 1/2^-[301]$ (the Nilsson notation) originating from the proton $p_{1/2}$ orbital. The very different structures between the $9/2^+$ ground state and the $1/2^-$ first excited state can  explain the fact that the $1/2^-$ state is a well-known isomer with a half-life of about 16 yr. Such a low-lying state could be populated readily at low stellar temperature $T \sim 1$ GK (corresponding to about 90 keV), and may potentially affect the stellar EC rates of $^{93}$Nb even at low temperatures. In Fig. \ref{fig:one}, the energy separation between this $1/2^-$ and the higher-lying states are also reproduced successfully. The implication of the existence of this energy gap in $^{93}$Nb is that contributions to the EC rates from the excited states could be suddenly enhanced at the stellar temperature $T \approx 6$ GK (corresponding roughly to 550 keV). The effect will be discussed later in Fig. \ref{fig:four}.

\begin{figure*}[htbp]
\begin{center}
  \includegraphics[width=0.80\textwidth]{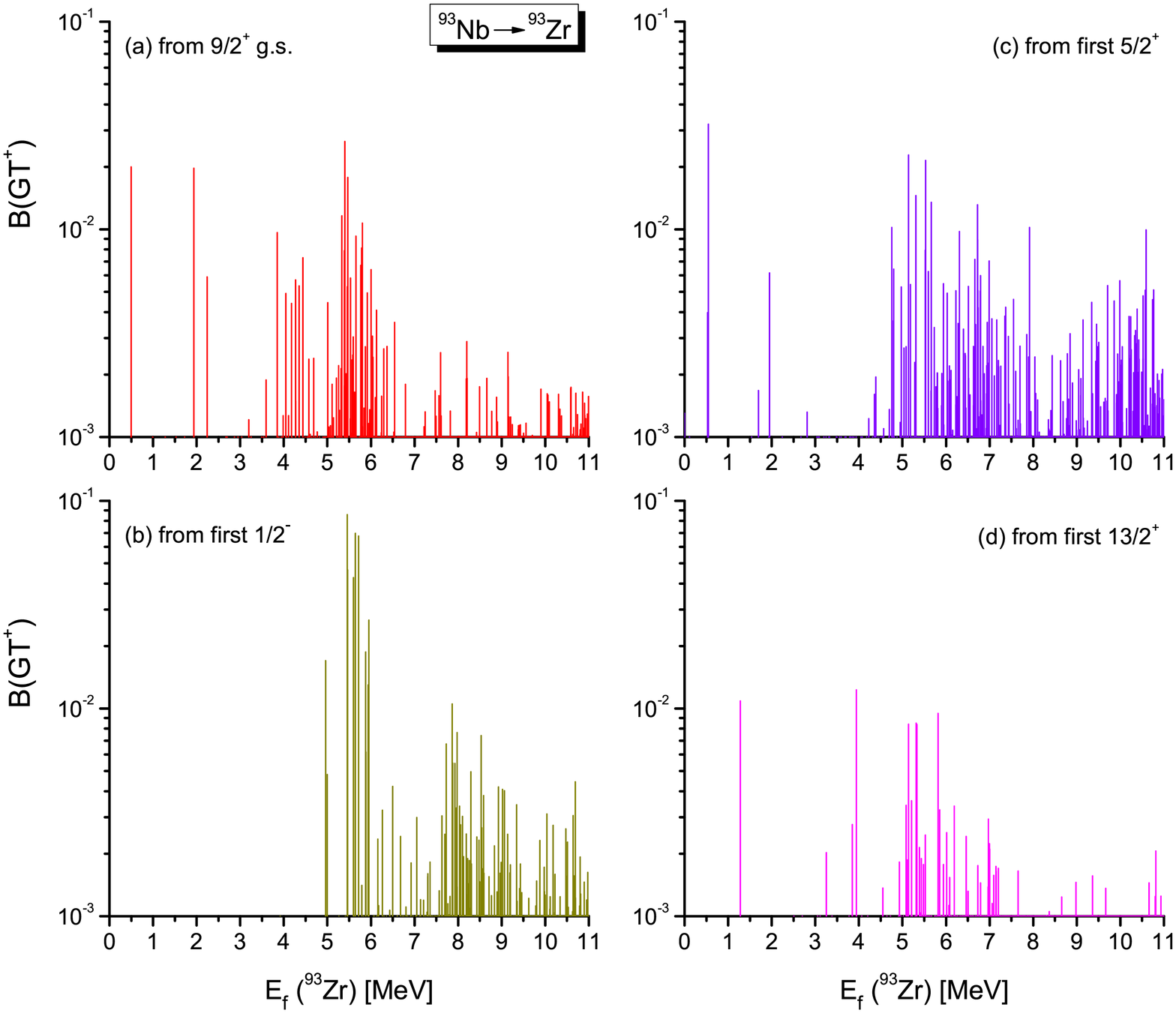}
  \caption{\label{fig:new} (Color online) The individual GT strength distribution $B(\mathrm{GT}^+)$ for the transitions from different states of the EC parent nucleus $^{93}$Nb to states of the daughter nucleus $^{93}$Zr as a function of its excitation energies, as calculated by the PSM. }
\end{center}
\end{figure*}

The EC daughter nucleus $^{93}$Zr is described by the present calculation only at a qualitative level, as shown in Fig. \ref{fig:one}. The energy of the experimental ground state $5/2^+$ is calculated to be higher than the first excited state $3/2^+$, although the energy gap between the $3/2^+$ and the higher-lying excited states is roughly reproduced. For this odd neutron nucleus with $N=53$, the last neutron occupies the $\{ d_{5/2}, g_{7/2}, s_{1/2} \}$ spherical mean-field orbitals, which can not be ideally described by models based on deformed mean fields such as the PSM. As we shall discuss in the next section, the relatively poor description of $^{93}$Zr will directly affect the GT strength distribution as differences in excitation energies of the daughter nucleus $^{93}$Zr will shift GT strength distribution peaks.

\subsection{GT strength distribution}


In Fig. \ref{fig:two}(a), we present our calculated GT strength distribution $B(\mathrm{GT}^+)$ for the transitions from the ground state $9/2^+$ of $^{93}$Nb to all obtained $\{7/2^+, 9/2^+, 11/2^+\}$ states of $^{93}$Zr as functions of the excitation energy $E_f$ (see red solid curve in Fig. \ref{fig:two}(a)). Our results are compared with the recent experimental data (black dots with the shaded area indicating statistical errors) \cite{EC_93Nb_PRC2020}, and with the calculations from the LSSM and QRPA. In the comparison with experiment, all the theoretical results are plotted in 0.5-MeV-wide bins as the data analysis does. 

For descriptions of the medium-heavy nuclei $^{93}$Nb and $^{93}$Zr, the cross-shell correlations in a large model space and multi-particle-multi-hole (multi-qp) correlations in large configuration space are expected to be important. In the LSSM calculation, a small model space with the $\{ 0f_{5/2}, 1p_{3/2}, 1p_{1/2}, 0g_{9/2} \}$ orbitals for protons and the $\{ 0g_{7/2}, 1d_{5/2}, 1d_{3/2}, 2s_{1/2}, 0h_{11/2} \}$ orbitals for neutrons is adopted, assuming a $^{78}$Ni core \cite{EC_93Nb_PRC2020}. To make the calculation feasible, the configuration space is further truncated to allow only up to three protons in the $0g_{9/2}$ orbital and no neutrons in the $0h_{11/2}$ orbital. To remedy these truncations, the calculated GT strengths $B(\mathrm{GT}^+)$ were renormalized (shrunk) by a phenomenological hindrance factor $h=5.43$ \cite{EC_93Nb_PRC2020}. It is seen from Fig. \ref{fig:two} (a) that with the above calculation conditions, most of the calculated $B(\mathrm{GT}^+)$ by the LSSM concentrate in two peaks roughly at $E_f \approx 2.5$ MeV and 4 MeV. No $B(\mathrm{GT}^+)$ strengths appear for $E_f \gtrsim 6$ MeV and the experimentally-observed fragmentation at higher excitations is not reproduced, which can be clearly seen in Fig. \ref{fig:two} (b) where the cumulative (running) sums of $B(\mathrm{GT}^+)$ are plotted.

For the QRPA, a larger model space is adopted for calculation, with the quenching factor $f_{\text{quench}} \approx 0.8$ \cite{EC_93Nb_PRC2020}. However, its configuration space is constructed with phonon-like structures that are summed by individual quasiparticle states. This is probably the reason that leads to a resonance-like GT peak at $E_f \approx 2.5$ MeV, which almost exhausts the $B(\mathrm{GT}^+)$ at the low-energy region, leaving little strengths for necessary fragmentations, as illustrated in  Fig. \ref{fig:two} (a) and (b). 

By comparison, the PSM employs a large model space (with four major harmonic-oscillator shells for both neutrons and protons) and a large configuration space (with up to 7-qp states with the configurations listed in Eqs. (\ref{Eq.config-n}, \ref{Eq.config-p})). For the present calculation from $^{93}$Nb's $9/2^+$ ground-state transition, for example, we include about 1000 daughter states for each of the $\{ 7/2^+, 9/2^+, 11/2^+ \}$ states in $^{93}$Zr, which is almost doubled as compared with the number of states in the LSSM calculation \cite{EC_93Nb_PRC2020}. The very distributed daughter states enable a more realistic description of the broad fragmentation of the experimental $B(\mathrm{GT}^+)$ strengths for the entire excitation-energy region. The fragmented GT distribution can be realized in Fig. \ref{fig:two}(a), with the four GT peaks observed in experiment are qualitatively reproduced by the PSM calculation. We note that the predicted locations for the first two peaks are shifted to higher energies, which is likely attributed to the problem in the description of level structure in $^{93}$Zr, as discussed in Fig. \ref{fig:one}. We also note that each of our 0.5-MeV-wide bin in Fig. \ref{fig:two}(a) may contain many individual transitions to the  daughter states. In Fig. \ref{fig:new}, we plot the original $B(\mathrm{GT}^+)$ values from the PSM calculation without being grouped in bins. The plot in Fig. \ref{fig:new} (a) indicates the calculated GT transitions from the ground state of $^{93}$Nb. Apart from the few strong strengths at low energies, at about 4 MeV and 5-6 MeV we find concentrations, corresponding to the $B(\mathrm{GT}^+)$ peaks in Fig. \ref{fig:two} (a). From 8 MeV up, many relatively weak strengths exist, which may bring important contribution to the behavior of the stellar EC at high temperatures, as we shall discuss later. 

We emphasize an important feature from the PSM results, which is in a clear distinction from the LSSM and QRPA, that from $E_f \gtrsim 8$ MeV, the PSM predicts an increasing trend for $\sum B(\mathrm{GT}^+)$ where the other calculations suggest vanishing strengths. This increasing trend in $\sum B(\mathrm{GT}^+)$ predicted by the PSM can be attributed to the contribution from the higher-order of qp configurations. Although individual strengths of the allowed GT transitions from the parent ground state to these states may not be significant, the number of such higher-order of qp configurations is large and increases with the excitation energy, and therefore, the total contribution may not be small. At this excitation energy and beyond where the uncertainty of experimental $B(\mathrm{GT}^+)$ gets larger due to the difficulty to separate the GT strengths from the isovector spin giant monopole resonance, the current data analysis stopped  \cite{EC_93Nb_PRC2020}.  

With growing excitation energies, the cumulative sum of $B(\mathrm{GT}^+)$ can be viewed in \ref{fig:two}(b). Among the theories, the LSSM (with renormalizations for the results) gives the best comparison with data up to about 6 MeV in excitation, and afterwards, begins to departure from the data curve because no new strengths further contribute. The discrepancy for the LSSM in the high-excitation region is clearly due to the lack of cross-shell correlations in its limited model space. The QRPA produces too large $\sum B(\mathrm{GT}^+)$ at the low-energy region but too little at the high-energy region. The reason leading to this result is clearly due to the resonance-like behavior at 2.5 MeV in the QRPA. The PSM gives qualitatively correct trend with increasing $\sum B(\mathrm{GT}^+)$, but the problem for the PSM result is that the increase is delayed to high energies due to inappropriate distribution of the energy levels in $^{93}$Zr, as discussed before. At 10 MeV, the PSM and QRPA calculations almost matches the experimental result, but begin to show an important distinction: The QRPA curve seems to keep flat while a robustly increasing trend starting from 6 MeV is predicted by the PSM. This difference occurs because of the applied algorithms for building configuration space in models. As the energy goes up, higher-order qp states in the PSM start to contribute to $\sum B(\mathrm{GT}^+)$ until the GT sumrule is reached, while no such mechanism exists in the QRPA. Note that 5-qp (7-qp) configurations for odd-mass nuclei in the PSM calculation correspond conceptually to the 2-particle-2-hole (3-particle-3-hole) excitations in the language of the QRPA.   

In stellar environments with finite temperatures, nuclei can have considerable probability to be thermally populated in excited nuclear states, which would in principle contribute to stellar EC rates. The contributions depend on strengths of the GT strength distributions as well as the precise excitation energies where the GT strengths locate. If no experimental data are available and no explicit theoretical calculations are performed, one assumes the Brink-Axel hypothesis. In Fig. \ref{fig:two} (c) and (d) we show respectively the $B(\mathrm{GT}^+)$ distributions and their cumulative sums for the transitions from three representative excited states (with different excitation energies and spin-parity values) of the parent nucleus $^{93}$Nb. For the very low-lying first $1/2^-$ state with experimental value at 30.77 keV, the calculated $B(\mathrm{GT}^+)$ is found to be negligible at $E_f \lesssim 5.0$ MeV (seen more clearly in Fig. \ref{fig:new}(b)). This is because that the $1/2^-$ state has a relatively pure wave function with the main configuration as $\pi 1/2^-[301]$ from the proton $p_{1/2}$ orbital, so that the connection by the GT operator with negative-parity states of $^{93}$Zr can only be the type that mixes strongly with the neutron $p_{1/2}$ orbital. Such states lie in higher energies across the neutron $N=50$ shell gap. Figure \ref{fig:new}(b) illustrates the situation. 

On the other hand, the wave function of the first $5/2^+$ state in $^{93}$Nb (see Fig. \ref{fig:one}) contains a strong mixture from many 1-qp and 3-qp configurations that involves the mean-field levels originating from the proton $g_{9/2}$ orbital. This orbital can be readily connected by the GT operator either with the neutron $g_{7/2}$ orbital at small energies, or with the neutron $g_{9/2}$ orbital at large energies crossing the $N=50$ shell gap. The $B(\mathrm{GT}^+)$ is then found to be sizable in both lower-lying ($E_f \approx 1$ MeV) and higher-lying ($E_f > 4$ MeV) regions (see Fig. \ref{fig:new}(c)), with a broad fragmentation due to the latter configurations, as shown in Fig. \ref{fig:two}(c). 

The wave function of the first high-spin $13/2^+$ state shows qualitatively a similar configuration mixing as in the first $5/2^+$ state. The difference is that the former tends to mix configurations that involve the high-$K$ components of the proton $g_{9/2}$ orbital, i.e. $\pi 9/2^+ [404]$ and $\pi 7/2^+ [413]$, while the latter mixes configurations with low-$K$ components. Concretely, the wave function of this $13/2^+$ state consists of $\sim 16\%$ of $\pi 1/2^+[440]$, $\sim 13\%$ of $\pi 3/2^+[431]$, $\sim 10\%$ of $\pi 5/2^+[422]$, $\sim 11\%$ of $\pi 3/2^+[431] \otimes 5/2^+[422] \otimes 5/2^+[422]$ with the $K=3/2$ configuration, and many other 3-qp and 5-qp configurations (some of which includes the $\pi 7/2[413]$ 1-qp configuration). The $B(\mathrm{GT}^+)$ from the first $13/2^+$ state is found to be much smaller in strength than the first $5/2^+$ state in the entire excitation-energy region, as seen from Fig. \ref{fig:two} (d) as well as Fig. \ref{fig:new} (c) and (d).

\subsection{Stellar electron-capture rates}

\begin{figure*}[htbp]
\begin{center}
  \includegraphics[width=0.80\textwidth]{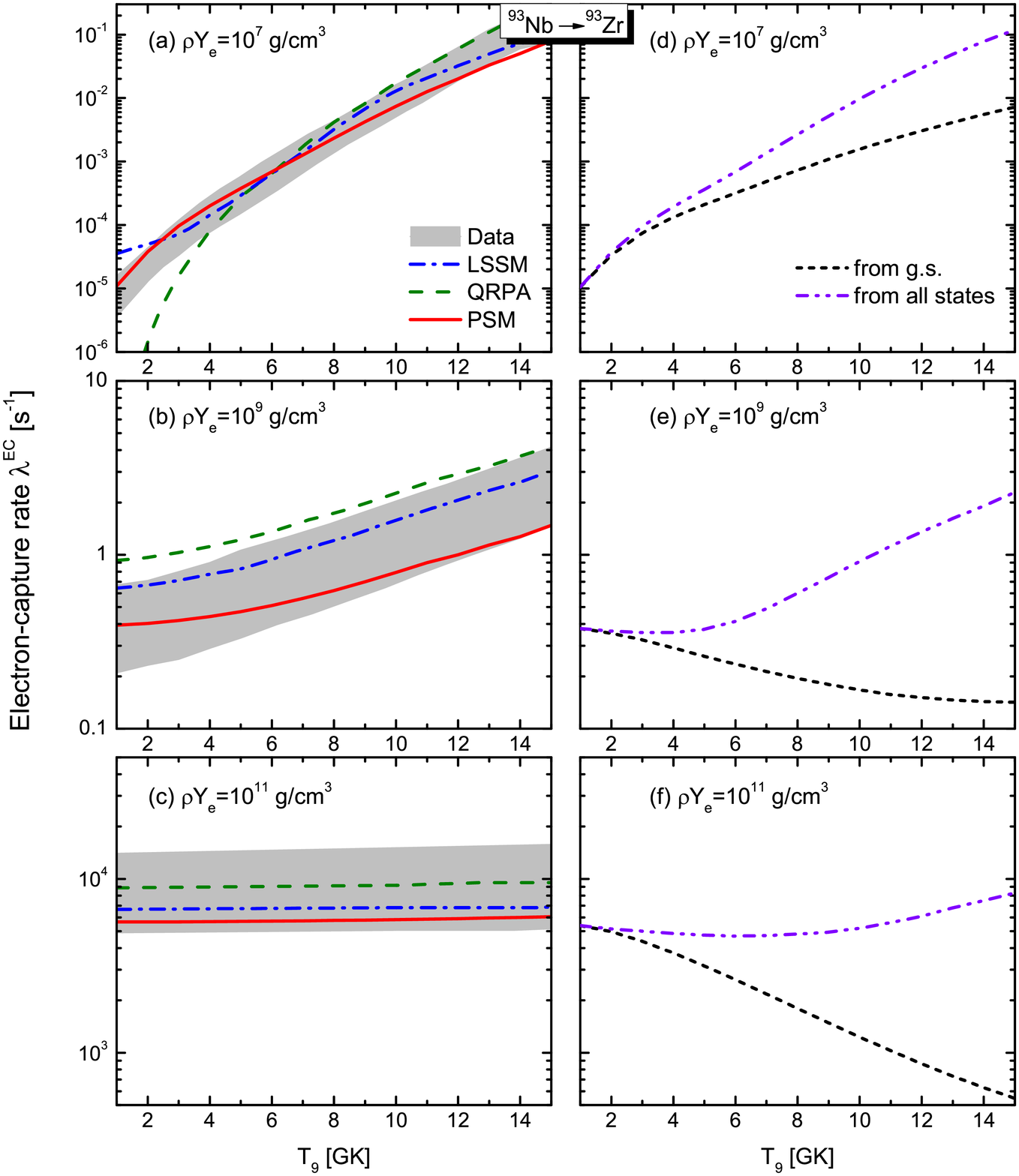}
  \caption{\label{fig:four} (Color online) The electron-capture rates for $^{93}$Nb $\rightarrow$ $^{93}$Zr as a function of the temperature $T_9$ (GK) at different stellar densities $\rho Y_e = 10^7, 10^9$ and $10^{11}$ g/cm$^3$. For the case where the parent nucleus $^{93}$Nb is supposed to stay in the ground state, calculations from three nuclear models are compared with each other and with the recent data [11] in panels (a-c), with the common legend indicated in (a). For the realistic case that all low-lying states of $^{93}$Nb could be populated in the finite-temperature condition, calculations of the projected shell model with and without the excited states are shown in panels (d-f), with the common legend indicated in (d). See the text for details.}
\end{center}
\end{figure*}

\begin{figure}[htbp]
\begin{center}
  \includegraphics[width=0.49\textwidth]{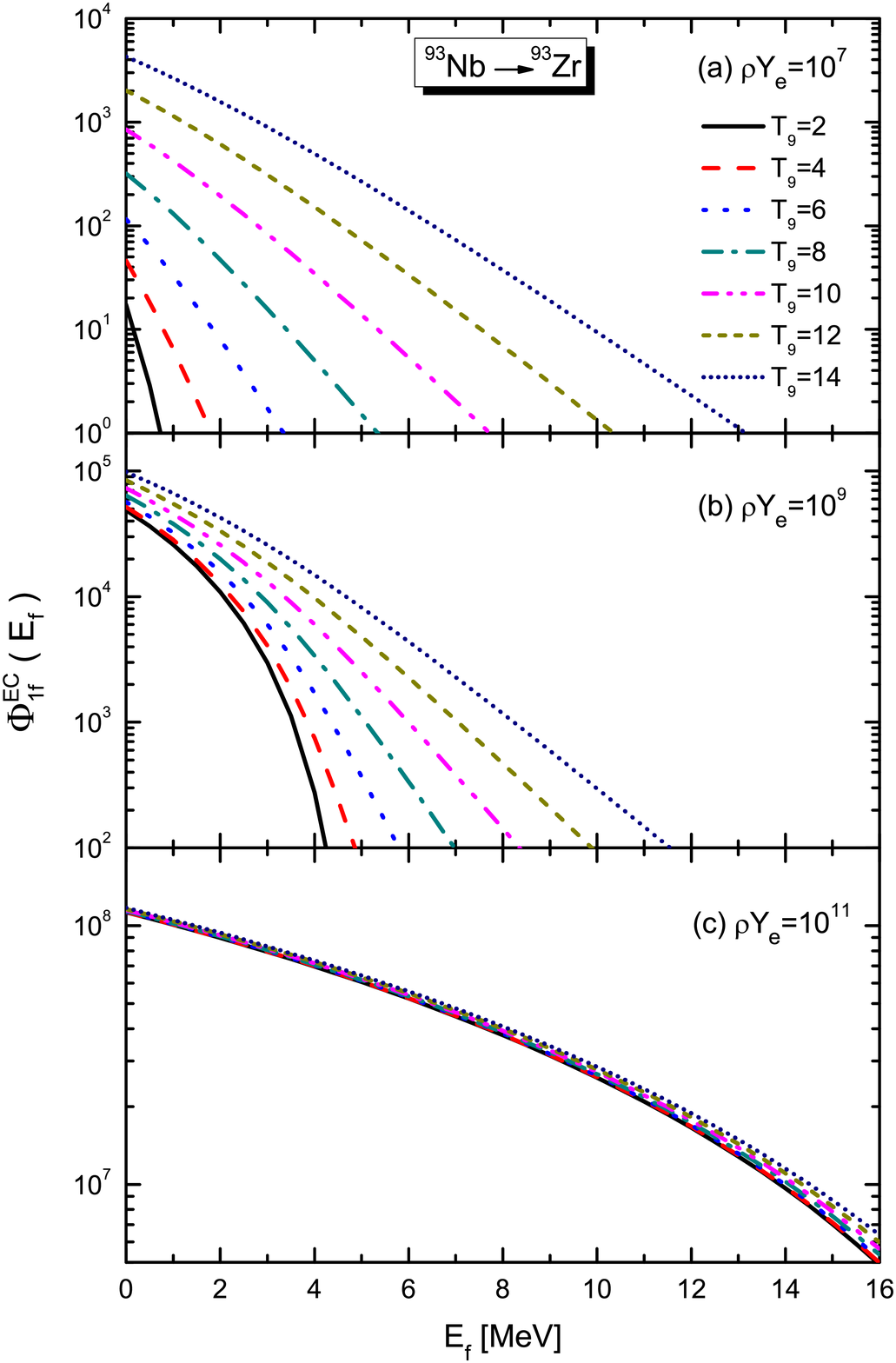}
  \caption{\label{fig:five} (Color online) The phase space integral $\Phi^{\text{EC}}_{1f}$ for transitions from the ground state of $^{93}$Nb to states of $^{93}$Zr at different stellar densities and temperatures, as a function of the excitation energy $E_f$. Note that the figure is in logarithmic scale and the scale is different in each panel. }
\end{center}
\end{figure}

With the calculated GT strength distributions, we can now study the  $^{93}$Nb stellar EC rates at different densities and temperatures. From the second summation over $f$ in Eq. (\ref{eq.lambda-EC}), it becomes evident that in order for a daughter state labeled $f$ to contribute significantly to the total $\lambda^{\text{EC}}$, it is required that the $f$-state necessarily has a large $B_{if}$. At the same time, the phase space must be open for the $f$-state to contribute, which requires a sizeable $\Phi_{if}^{\rm EC}$.  

We first discuss a popular treatment for electron capture processes occurring in astrophysical problems. In such a simplified treatment, the parent nucleus $^{93}$Nb is supposed to stay in its ground state all the time regardless of thermal excitations. This simplification thus completely neglects detailed thermal populations of the parent-nucleus states, and corresponds to replacing the $i$-summation in Eq. (\ref{eq.lambda-EC}) by a unity and setting $i \equiv 1$ in the second summation. We may call this treatment zero-temperature approximation for parent nuclei. The resulting EC rates as functions of the temperature $T_9$ (GK) at three typical stellar densities are shown in Fig. \ref{fig:four} (a), (b) and (c). Among the three density conditions,  $\rho Y_e = 10^7$ g/cm$^3$ corresponds to the condition of the core during the silicon-burning stage in a pre-supernova star, $\rho Y_e = 10^9$ g/cm$^3$ represents the condition for the pre-collapse of the core for massive stars, and $\rho Y_e = 10^{11}$ g/cm$^3$ labels the onset of the core collapse. Calculations from the three different nuclear models are compared with each other, and with the recent experimental values \cite{EC_93Nb_PRC2020}. 

As the ground-state to ground-state EC $Q$-value is as small as $Q_{11} = -0.602$ MeV in the present example, the contribution to stellar EC rates from transitions to excited states of the daughter nucleus $^{93}$Zr would be considerable even at low stellar densities and temperatures. For the low-density case with $\rho Y_e = 10^7$ g/cm$^3$, the electron chemical potential $\mu_e \approx 1$ MeV. From Eq. (\ref{eq.lambda-EC}) and the corresponding phase space integral $\Phi^{\text{EC}}_{1f}$ distributions in Fig. \ref{fig:five} (a), at low temperatures of $T_9 \lesssim 4$ GK, transitions to the  low-lying states of $^{93}$Zr with $E_f \lesssim 1$ MeV are important. As the QRPA shows no $B(\mathrm{GT}^+)$ strength in such low-excitation region (see Fig. \ref{fig:two}(b)), the calculated EC rates are strongly underestimated as seen from Fig. \ref{fig:four}(a). With  increasing temperature, the electron Fermi surface becomes more diffuse and the corresponding phase space integral $\Phi^{\text{EC}}_{1f}$ increases rapidly as seen from Fig. \ref{fig:five} (a). The transitions to higher excited states of $^{93}$Zr within $1 \lesssim E_f \lesssim 2$ MeV begin to contribute. Within experimental uncertainties, the derived EC rates from the three nuclear models (LSSM, QRPA, and PSM) are all consistent with data for $T_9 > 4$ GK. 

For the $\rho Y_e = 10^9$ g/cm$^3$ case, the electron chemical potential increases to $\mu_e \approx 5$ MeV. The contributions of GT transitions to the excited states of $^{93}$Zr with $E_f \lesssim 5.5$ MeV would play a role in the stellar EC rates. From the $\Phi^{\text{EC}}_{1f}$ distributions in Fig. \ref{fig:five} (b), the states with higher excitations in $^{93}$Zr would be involved effectively when $T_9$ goes high. The EC rates are found not as sensitive to the details of the GT strength distribution as in the low-density case, but largely depend on both the GT strength distribution and the integrated GT strength for $E_f \lesssim 5.5$ MeV. As a consequence, differences in calculated results from different models can be as large as several times. It was seen from Fig. \ref{fig:two}(b) that for this energy region, the QRPA overestimates the  $B(\mathrm{GT}^+)$, the PSM underestimates that, and the SM (with renormalizations for the GT results) agrees better with the data. The derived EC rates by the SM and PSM are compared well with data within the uncertainties, but for the QRPA, the calculation is a little bit overestimated, as seen from Fig. \ref{fig:four}(b). Unfortunately, due to large uncertainties, the current experimental data are not able to discriminate the models.

For the high density case with $\rho Y_e = 10^{11}$ g/cm$^3$, the electron chemical potential increases to about 20 MeV. As seen from Fig. \ref{fig:five}(c), the phase space for this density condition opens for all excited states that are included in the plot, with $\Phi^{\text{EC}}_{1f}$ (nearly) exponentially decreasing with $E_f$ and essentially being independent of stellar temperatures. The derived stellar EC rates then mainly depends on the integrated GT strength. As seen from Fig. \ref{fig:four}(c), the calculated $\lambda^{\text{EC}}$ by all the three models fall consistently within data uncertainties, with no temperature-dependence. One should keep in mind that these $\lambda^{\text{EC}}$ results are obtained with the zero-temperature approximation, i.e. when the assumption that electrons are captured always by the ground state of the parent $^{93}$Nb is applied.

When the $i$-summation for the parent-nucleus states is fully taken into account in the stellar EC rates expression in Eq. (\ref{eq.lambda-EC}),  thermal populations of different nuclear states may vary considerably with stellar temperatures, and we thus call it the finite-temperature environment. This can be clearly seen in Eq. (\ref{eq.lambda-EC}) through the $i$-summation content $ \sum_i (2J_i+1) e^{-E_i/(k_BT)} / G(Z, A, T) $, which depend explicitly on $E_i$, $J_i$ and $T$. In general, as temperature goes up, excited states are successively populated while the occupation probability of the ground state drops down. To see clearly how population variations in the nuclear states affect the EC rates, we present in Fig. \ref{fig:four} (d-f) the PSM calculations for EC rates with and without the contributions of excited states of $^{93}$Nb. When only transitions from the $9/2^+$ ground state of $^{93}$Nb are considered in the finite-temperature environment (i.e. we extract only the $i=1$ term from the summation),  the occupation probability of the ground state decreases with temperature, and the corresponding EC rates are smaller when compared with the calculations in Fig. \ref{fig:four} (a-c) where the parent nucleus is supposed to stay in its ground state all the time. For the $\rho Y_e = 10^7$ g/cm$^3$ case with small $\mu_e$, EC rates climb rapidly with temperature and the effect of the thermal occupation is not very significant if we compare Fig. \ref{fig:four} (d) with (a). For the $\rho Y_e = 10^9$ g/cm$^3$ case with large $\mu_e$, the EC rates in Fig. \ref{fig:four} (b) only show moderate increase with the temperature. The effect of the thermal occupation of the ground state leads to a clear decrease of the EC rates with temperature, if we  compare the black-dotted curve in Fig. \ref{fig:four} (e) and the PSM results in Fig. \ref{fig:four} (b). Similar large differences can also be seen when comparing Fig. \ref{fig:four} (f) and (c).

With all the excited states of the parent nucleus $^{93}$Nb fully taken into account in the $i$-summation in Eq. (\ref{eq.lambda-EC}), one can see how the EC rates are changed by comparing the two curves in Fig. \ref{fig:four} (d-f). At low temperatures with $T_9 \lesssim 3$ GK, only the first excited $1/2^-$ state in $^{93}$Nb is populated effectively. Since this state has negligible $B(\mathrm{GT}^+)$ at $E_f \lesssim 5.5$ MeV (see Fig. \ref{fig:two}(d)) and a small $2J+1$ factor (see Eq. (\ref{eq.lambda-EC})), the effect of thermal population of excited parent-nucleus states is found to be very small (namely, the two curves in Fig. \ref{fig:four} (d-f) in this temperature region are very close to each other). This conclusion has been obtained in a recent simulation by Misch {\it et al.} \cite{Wendell_2021_tmp}. At higher temperatures with $T_9 \gtrsim 6$ GK (corresponding roughly to 550 keV), many excited states with sizable $2J+1$ factors begin to be populated effectively (see Fig. \ref{fig:one}). The $B(\mathrm{GT}^+)$ strengths of these excited states are comparable to that of the ground state (see Fig. \ref{fig:two}), and the effects on EC rates are thus significant. The realistic stellar EC rates (and the effects of excited parent-nucleus states) are found to increase rapidly with temperature. At $T_9 \approx 15$ GK, the effect of excited parent-nucleus states is found to be more than one order of magnitude when comparing the two curves in Fig. \ref{fig:four} (d-f). 

Finally, it is illustrative to compare the two calculated EC rates for $^{93}$Nb from the PSM: those in the zero-temperature approximation in Fig. \ref{fig:four}(a-c) (red solid lines) and those in the full calculations in Fig. \ref{fig:four}(d-f) (labeled as `from all states'). It is seen that similar curves are obtained in the two cases, except when the stellar temperatures are very high (with $T_9 \gtrsim 10$ GK) the full calculation gives slightly higher EC rates (which is illustrated clearly in Fig. \ref{fig:four} (c) and (f)). This may suggest that the zero-temperature approximation used for stellar EC rates is reasonably good. However, we caution that this conclusion should be taken as special for the present $^{93}$Nb $\rightarrow$ $^{93}$Zr case because for the EC parent nucleus $^{93}$Nb, the very-low first $1/2^-$ state has negligible $B(\mathrm{GT}^+)$ values, and furthermore, there exists a large energy gap between this $1/2^-$ state and the other excited states, which diminishes their contributions to the EC rates. 

\section{\label{sec:sum}summary}

To summarize, electron capture (EC) rates are important nuclear inputs for understanding many astrophysical phenomena such as the core-collapse supernovae, the Urca cooling of neutron star crust, etc. In such stellar environments with high temperature and density, the contribution of excited nuclear states to the EC rates is indispensable, and their effects should be considered. Astrophysical simulations usually take one of the two approaches to the distribution of nuclear states: either they assume a thermal equilibrium distribution (e.g. a finite-temperature environment with the Boltzmann distribution), or they use only the ground state properties (the zero-temperature approximation). With new development of nuclear many-body techniques, explicit treatment of all excited states in theoretical calculations become possible. This allows one to compare the novel method with the traditional ones before it can be applied in astrophysical simulations. 

In the present article, we introduced a Projected-Shell-Model calculation for stellar EC rates in medium-heavy odd-mass nuclei by taking $^{93}$Nb $\rightarrow$ $^{93}$Zr as the example. In the zero-temperature approximation with only the ground state of the parent nucleus considered, the calculated Gamow-Teller transition strength distribution $B(\mathrm{GT}^+)$ were compared with the recent experimental data \cite{EC_93Nb_PRC2020} and discussed together with the large-scale shell model calculation and the quasiparticle random-phase approximation. As seen in Fig. \ref{fig:two} (a) and (b), theoretical $B(\mathrm{GT}^+)$ from different models give large deviations from each other, showing even no qualitative similarities. We have discussed possible reasons that lead to the characteristic GT pattern for each model. We further discussed the derived stellar EC rates at different temperatures and densities. We found that for the present example, although different nuclear models give very different GT strength distributions, the derived stellar EC rates are all qualitatively consistent with the data within the statistical errors and uncertainties for most cases.
In the full calculations by the Projected Shell Model with all excited parent-nucleus states taking into account exactly, similar EC rates are obtained as compared to those from the zero-temperature approximation in which only the ground state properties of the parent nucleus is considered. This seems to suggest equivalence of the two approaches to including the effects of excited nuclear states. 

The insensitivity of the stellar EC rates found in the present work, to detailed nuclear structure and to different approaches to treat the contribution of nuclear states, needs further investigation. In the present $^{93}$Nb case, the electron-capture $Q$ value is close to zero, and when the Fermi energy is sufficiently high, the details of the GT strength distribution do not have a strong impact on the derived EC rates. We however caution that this conclusion for $^{93}$Nb should not be taken as general since in principle the EC rates are expected to depend on detailed nuclear-structure properties such as the energy levels, $B(\mathrm{GT}^+)$, $Q$-values, and phase space factors, etc. For example, the EC rates of $^{59}$Co in the full calculations are by one order of magnitude larger than those from the zero-temperature approximation in most temperature-density conditions \cite{Tan_2020_PLB}. Experimental data with improved resolution, which can   discriminate theoretical models, are very much desired. 

\begin{acknowledgments}
  This work is supported by the Fundamental Research Funds for the Central Universities (with Grant No. SWU019013), by the National Natural Science Foundation of China (Grant Nos. 11905175, 11875225, and U1932206), by the National Key Program for S$\&$T Research and Development (Grant No. 2016YFA0400501), and by the Venture $\&$ Innovation Support Program for Chongqing Overseas Returnees (with Grant No. cx2019056).
\end{acknowledgments}



\begin{thebibliography}{58}%
\makeatletter
\providecommand \@ifxundefined [1]{%
 \@ifx{#1\undefined}
}%
\providecommand \@ifnum [1]{%
 \ifnum #1\expandafter \@firstoftwo
 \else \expandafter \@secondoftwo
 \fi
}%
\providecommand \@ifx [1]{%
 \ifx #1\expandafter \@firstoftwo
 \else \expandafter \@secondoftwo
 \fi
}%
\providecommand \natexlab [1]{#1}%
\providecommand \enquote  [1]{``#1''}%
\providecommand \bibnamefont  [1]{#1}%
\providecommand \bibfnamefont [1]{#1}%
\providecommand \citenamefont [1]{#1}%
\providecommand \href@noop [0]{\@secondoftwo}%
\providecommand \href [0]{\begingroup \@sanitize@url \@href}%
\providecommand \@href[1]{\@@startlink{#1}\@@href}%
\providecommand \@@href[1]{\endgroup#1\@@endlink}%
\providecommand \@sanitize@url [0]{\catcode `\\12\catcode `\$12\catcode
  `\&12\catcode `\#12\catcode `\^12\catcode `\_12\catcode `\%12\relax}%
\providecommand \@@startlink[1]{}%
\providecommand \@@endlink[0]{}%
\providecommand \url  [0]{\begingroup\@sanitize@url \@url }%
\providecommand \@url [1]{\endgroup\@href {#1}{\urlprefix }}%
\providecommand \urlprefix  [0]{URL }%
\providecommand \Eprint [0]{\href }%
\providecommand \doibase [0]{https://doi.org/}%
\providecommand \selectlanguage [0]{\@gobble}%
\providecommand \bibinfo  [0]{\@secondoftwo}%
\providecommand \bibfield  [0]{\@secondoftwo}%
\providecommand \translation [1]{[#1]}%
\providecommand \BibitemOpen [0]{}%
\providecommand \bibitemStop [0]{}%
\providecommand \bibitemNoStop [0]{.\EOS\space}%
\providecommand \EOS [0]{\spacefactor3000\relax}%
\providecommand \BibitemShut  [1]{\csname bibitem#1\endcsname}%
\let\auto@bib@innerbib\@empty
\bibitem [{\citenamefont {Fuller}\ \emph {et~al.}(1980)\citenamefont {Fuller},
  \citenamefont {Fowler},\ and\ \citenamefont {Newman}}]{Fuller1}%
  \BibitemOpen
  \bibfield  {author} {\bibinfo {author} {\bibfnamefont {G.~M.}\ \bibnamefont
  {Fuller}}, \bibinfo {author} {\bibfnamefont {W.~A.}\ \bibnamefont {Fowler}},\
  and\ \bibinfo {author} {\bibfnamefont {M.~J.}\ \bibnamefont {Newman}},\
  }\bibfield  {title} {\bibinfo {title} {Stellar weak-interaction rates for
  $sd$-shell nuclei. $\text{I}$. nuclear matrix element systematics with
  application to $^{26}\text{Al}$ and selected nuclei of imprtance to the
  supernova problem},\ }\href@noop {} {\bibfield  {journal} {\bibinfo
  {journal} {Astrophys. J. Suppl.}\ }\textbf {\bibinfo {volume} {42}},\
  \bibinfo {pages} {447} (\bibinfo {year} {1980})}\BibitemShut {NoStop}%
\bibitem [{\citenamefont {Fuller}\ \emph
  {et~al.}(1982{\natexlab{a}})\citenamefont {Fuller}, \citenamefont {Fowler},\
  and\ \citenamefont {Newman}}]{Fuller2}%
  \BibitemOpen
  \bibfield  {author} {\bibinfo {author} {\bibfnamefont {G.~M.}\ \bibnamefont
  {Fuller}}, \bibinfo {author} {\bibfnamefont {W.~A.}\ \bibnamefont {Fowler}},\
  and\ \bibinfo {author} {\bibfnamefont {M.~J.}\ \bibnamefont {Newman}},\
  }\bibfield  {title} {\bibinfo {title} {Stellar weak-interaction rates for
  intermediate-mass nuclei. $\text{II}$.},\ }\href@noop {} {\bibfield
  {journal} {\bibinfo  {journal} {Astrophys. J.}\ }\textbf {\bibinfo {volume}
  {252}},\ \bibinfo {pages} {715} (\bibinfo {year}
  {1982}{\natexlab{a}})}\BibitemShut {NoStop}%
\bibitem [{\citenamefont {Fuller}\ \emph
  {et~al.}(1982{\natexlab{b}})\citenamefont {Fuller}, \citenamefont {Fowler},\
  and\ \citenamefont {Newman}}]{Fuller3}%
  \BibitemOpen
  \bibfield  {author} {\bibinfo {author} {\bibfnamefont {G.~M.}\ \bibnamefont
  {Fuller}}, \bibinfo {author} {\bibfnamefont {W.~A.}\ \bibnamefont {Fowler}},\
  and\ \bibinfo {author} {\bibfnamefont {M.~J.}\ \bibnamefont {Newman}},\
  }\bibfield  {title} {\bibinfo {title} {Stellar weak-interaction rates for
  intermediate-mass nuclei. $\text{III}$.},\ }\href@noop {} {\bibfield
  {journal} {\bibinfo  {journal} {Astrophys. J. Suppl.}\ }\textbf {\bibinfo
  {volume} {48}},\ \bibinfo {pages} {279} (\bibinfo {year}
  {1982}{\natexlab{b}})}\BibitemShut {NoStop}%
\bibitem [{\citenamefont {Fuller}\ \emph {et~al.}(1985)\citenamefont {Fuller},
  \citenamefont {Fowler},\ and\ \citenamefont {Newman}}]{Fuller4}%
  \BibitemOpen
  \bibfield  {author} {\bibinfo {author} {\bibfnamefont {G.~M.}\ \bibnamefont
  {Fuller}}, \bibinfo {author} {\bibfnamefont {W.~A.}\ \bibnamefont {Fowler}},\
  and\ \bibinfo {author} {\bibfnamefont {M.~J.}\ \bibnamefont {Newman}},\
  }\bibfield  {title} {\bibinfo {title} {Stellar weak-interaction rates for
  intermediate-mass nuclei. ${IV}$.},\ }\href@noop {} {\bibfield  {journal}
  {\bibinfo  {journal} {Astrophys. J.}\ }\textbf {\bibinfo {volume} {293}},\
  \bibinfo {pages} {1} (\bibinfo {year} {1985})}\BibitemShut {NoStop}%
\bibitem [{\citenamefont {Heger}\ \emph {et~al.}(2001)\citenamefont {Heger},
  \citenamefont {Langanke}, \citenamefont {Mart\'{\i}nez-Pinedo},\ and\
  \citenamefont {Woosley}}]{Heger_2001_PRL}%
  \BibitemOpen
  \bibfield  {author} {\bibinfo {author} {\bibfnamefont {A.}~\bibnamefont
  {Heger}}, \bibinfo {author} {\bibfnamefont {K.}~\bibnamefont {Langanke}},
  \bibinfo {author} {\bibfnamefont {G.}~\bibnamefont {Mart\'{\i}nez-Pinedo}},\
  and\ \bibinfo {author} {\bibfnamefont {S.~E.}\ \bibnamefont {Woosley}},\
  }\bibfield  {title} {\bibinfo {title} {Presupernova collapse models with
  improved weak-interaction rates},\ }\href
  {https://doi.org/10.1103/PhysRevLett.86.1678} {\bibfield  {journal} {\bibinfo
   {journal} {Phys. Rev. Lett.}\ }\textbf {\bibinfo {volume} {86}},\ \bibinfo
  {pages} {1678} (\bibinfo {year} {2001})}\BibitemShut {NoStop}%
\bibitem [{\citenamefont {Hix}\ \emph {et~al.}(2003)\citenamefont {Hix},
  \citenamefont {Messer}, \citenamefont {Mezzacappa}, \citenamefont
  {Liebend\"orfer}, \citenamefont {Sampaio}, \citenamefont {Langanke},
  \citenamefont {Dean},\ and\ \citenamefont
  {Mart\'{\i}nez-Pinedo}}]{Hix_2003_PRL}%
  \BibitemOpen
  \bibfield  {author} {\bibinfo {author} {\bibfnamefont {W.~R.}\ \bibnamefont
  {Hix}}, \bibinfo {author} {\bibfnamefont {O.~E.~B.}\ \bibnamefont {Messer}},
  \bibinfo {author} {\bibfnamefont {A.}~\bibnamefont {Mezzacappa}}, \bibinfo
  {author} {\bibfnamefont {M.}~\bibnamefont {Liebend\"orfer}}, \bibinfo
  {author} {\bibfnamefont {J.}~\bibnamefont {Sampaio}}, \bibinfo {author}
  {\bibfnamefont {K.}~\bibnamefont {Langanke}}, \bibinfo {author}
  {\bibfnamefont {D.~J.}\ \bibnamefont {Dean}},\ and\ \bibinfo {author}
  {\bibfnamefont {G.}~\bibnamefont {Mart\'{\i}nez-Pinedo}},\ }\bibfield
  {title} {\bibinfo {title} {Consequences of nuclear electron capture in core
  collapse supernovae},\ }\href {https://doi.org/10.1103/PhysRevLett.91.201102}
  {\bibfield  {journal} {\bibinfo  {journal} {Phys. Rev. Lett.}\ }\textbf
  {\bibinfo {volume} {91}},\ \bibinfo {pages} {201102} (\bibinfo {year}
  {2003})}\BibitemShut {NoStop}%
\bibitem [{\citenamefont {Langanke}\ and\ \citenamefont
  {Mart\'{\i}nez-Pinedo}(2003)}]{Langanke-RMP}%
  \BibitemOpen
  \bibfield  {author} {\bibinfo {author} {\bibfnamefont {K.}~\bibnamefont
  {Langanke}}\ and\ \bibinfo {author} {\bibfnamefont {G.}~\bibnamefont
  {Mart\'{\i}nez-Pinedo}},\ }\bibfield  {title} {\bibinfo {title} {Nuclear
  weak-interaction processes in stars},\ }\href@noop {} {\bibfield  {journal}
  {\bibinfo  {journal} {Rev. Mod. Phys.}\ }\textbf {\bibinfo {volume} {75}},\
  \bibinfo {pages} {819} (\bibinfo {year} {2003})}\BibitemShut {NoStop}%
\bibitem [{\citenamefont {Langanke}\ \emph {et~al.}(2021)\citenamefont
  {Langanke}, \citenamefont {Mart{\'\i}nez-Pinedo},\ and\ \citenamefont
  {Zegers}}]{langanke_2020_EC_review}%
  \BibitemOpen
  \bibfield  {author} {\bibinfo {author} {\bibfnamefont {K.}~\bibnamefont
  {Langanke}}, \bibinfo {author} {\bibfnamefont {G.}~\bibnamefont
  {Mart{\'\i}nez-Pinedo}},\ and\ \bibinfo {author} {\bibfnamefont {R.~G.~T.}\
  \bibnamefont {Zegers}},\ }\bibfield  {title} {\bibinfo {title} {Electron
  capture in stars},\ }\href
  {http://iopscience.iop.org/article/10.1088/1361-6633/abf207} {\bibfield
  {journal} {\bibinfo  {journal} {Rep. Prog. Phys.}\ }\textbf {\bibinfo
  {volume} {84}},\ \bibinfo {pages} {066301} (\bibinfo {year}
  {2021})}\BibitemShut {NoStop}%
\bibitem [{\citenamefont {Janka}\ \emph {et~al.}(2007)\citenamefont {Janka},
  \citenamefont {Langanke}, \citenamefont {Marek}, \citenamefont
  {Mart\'{\i}nez-Pinedo},\ and\ \citenamefont {M\"uller}}]{physrep-2007}%
  \BibitemOpen
  \bibfield  {author} {\bibinfo {author} {\bibfnamefont {H.-T.}\ \bibnamefont
  {Janka}}, \bibinfo {author} {\bibfnamefont {K.}~\bibnamefont {Langanke}},
  \bibinfo {author} {\bibfnamefont {A.}~\bibnamefont {Marek}}, \bibinfo
  {author} {\bibfnamefont {G.}~\bibnamefont {Mart\'{\i}nez-Pinedo}},\ and\
  \bibinfo {author} {\bibfnamefont {B.}~\bibnamefont {M\"uller}},\ }\bibfield
  {title} {\bibinfo {title} {Theory of core-collapse supernovae},\ }\href@noop
  {} {\bibfield  {journal} {\bibinfo  {journal} {Phys. Rep.}\ }\textbf
  {\bibinfo {volume} {442}},\ \bibinfo {pages} {38} (\bibinfo {year}
  {2007})}\BibitemShut {NoStop}%
\bibitem [{\citenamefont {Cole}\ \emph {et~al.}(2012)\citenamefont {Cole},
  \citenamefont {Anderson}, \citenamefont {Zegers}, \citenamefont {Austin},
  \citenamefont {Brown}, \citenamefont {Valdez}, \citenamefont {Gupta},
  \citenamefont {Hitt},\ and\ \citenamefont {Fawwaz}}]{cole-2012}%
  \BibitemOpen
  \bibfield  {author} {\bibinfo {author} {\bibfnamefont {A.~L.}\ \bibnamefont
  {Cole}}, \bibinfo {author} {\bibfnamefont {T.~S.}\ \bibnamefont {Anderson}},
  \bibinfo {author} {\bibfnamefont {R.~G.~T.}\ \bibnamefont {Zegers}}, \bibinfo
  {author} {\bibfnamefont {S.~M.}\ \bibnamefont {Austin}}, \bibinfo {author}
  {\bibfnamefont {B.~A.}\ \bibnamefont {Brown}}, \bibinfo {author}
  {\bibfnamefont {L.}~\bibnamefont {Valdez}}, \bibinfo {author} {\bibfnamefont
  {S.}~\bibnamefont {Gupta}}, \bibinfo {author} {\bibfnamefont {G.~W.}\
  \bibnamefont {Hitt}},\ and\ \bibinfo {author} {\bibfnamefont
  {O.}~\bibnamefont {Fawwaz}},\ }\bibfield  {title} {\bibinfo {title}
  {Gamow-teller strengths and electron-capture rates for $pf$-shell nuclei of
  relevance for late stellar evolution},\ }\href@noop {} {\bibfield  {journal}
  {\bibinfo  {journal} {Phys. Rev. C}\ }\textbf {\bibinfo {volume} {86}},\
  \bibinfo {pages} {015809} (\bibinfo {year} {2012})}\BibitemShut {NoStop}%
\bibitem [{\citenamefont {Gao}\ \emph {et~al.}(2020)\citenamefont {Gao},
  \citenamefont {Zegers}, \citenamefont {Zamora}, \citenamefont {Bazin},
  \citenamefont {Brown}, \citenamefont {Bender}, \citenamefont {Crawford},
  \citenamefont {Engel}, \citenamefont {Falduto}, \citenamefont {Gade},
  \citenamefont {Gastis}, \citenamefont {Ginter}, \citenamefont {Guess},
  \citenamefont {Lipschutz}, \citenamefont {Macchiavelli}, \citenamefont
  {Miki}, \citenamefont {Ney}, \citenamefont {Longfellow}, \citenamefont
  {Noji}, \citenamefont {Pereira}, \citenamefont {Schmitt}, \citenamefont
  {Sullivan}, \citenamefont {Titus},\ and\ \citenamefont
  {Weisshaar}}]{EC_93Nb_PRC2020}%
  \BibitemOpen
  \bibfield  {author} {\bibinfo {author} {\bibfnamefont {B.}~\bibnamefont
  {Gao}}, \bibinfo {author} {\bibfnamefont {R.~G.~T.}\ \bibnamefont {Zegers}},
  \bibinfo {author} {\bibfnamefont {J.~C.}\ \bibnamefont {Zamora}}, \bibinfo
  {author} {\bibfnamefont {D.}~\bibnamefont {Bazin}}, \bibinfo {author}
  {\bibfnamefont {B.~A.}\ \bibnamefont {Brown}}, \bibinfo {author}
  {\bibfnamefont {P.}~\bibnamefont {Bender}}, \bibinfo {author} {\bibfnamefont
  {H.~L.}\ \bibnamefont {Crawford}}, \bibinfo {author} {\bibfnamefont
  {J.}~\bibnamefont {Engel}}, \bibinfo {author} {\bibfnamefont
  {A.}~\bibnamefont {Falduto}}, \bibinfo {author} {\bibfnamefont
  {A.}~\bibnamefont {Gade}}, \bibinfo {author} {\bibfnamefont {P.}~\bibnamefont
  {Gastis}}, \bibinfo {author} {\bibfnamefont {T.}~\bibnamefont {Ginter}},
  \bibinfo {author} {\bibfnamefont {C.~J.}\ \bibnamefont {Guess}}, \bibinfo
  {author} {\bibfnamefont {S.}~\bibnamefont {Lipschutz}}, \bibinfo {author}
  {\bibfnamefont {A.~O.}\ \bibnamefont {Macchiavelli}}, \bibinfo {author}
  {\bibfnamefont {K.}~\bibnamefont {Miki}}, \bibinfo {author} {\bibfnamefont
  {E.~M.}\ \bibnamefont {Ney}}, \bibinfo {author} {\bibfnamefont
  {B.}~\bibnamefont {Longfellow}}, \bibinfo {author} {\bibfnamefont
  {S.}~\bibnamefont {Noji}}, \bibinfo {author} {\bibfnamefont {J.}~\bibnamefont
  {Pereira}}, \bibinfo {author} {\bibfnamefont {J.}~\bibnamefont {Schmitt}},
  \bibinfo {author} {\bibfnamefont {C.}~\bibnamefont {Sullivan}}, \bibinfo
  {author} {\bibfnamefont {R.}~\bibnamefont {Titus}},\ and\ \bibinfo {author}
  {\bibfnamefont {D.}~\bibnamefont {Weisshaar}},\ }\bibfield  {title} {\bibinfo
  {title} {Gamow-teller transitions to $^{93}\mathrm{Zr}$ via the
  $^{93}\mathrm{Nb}(t,^{3}\mathrm{He}+\ensuremath{\gamma}$) reaction at 115
  mev/u and its application to the stellar electron-capture rates},\ }\href
  {https://doi.org/10.1103/PhysRevC.101.014308} {\bibfield  {journal} {\bibinfo
   {journal} {Phys. Rev. C}\ }\textbf {\bibinfo {volume} {101}},\ \bibinfo
  {pages} {014308} (\bibinfo {year} {2020})}\BibitemShut {NoStop}%
\bibitem [{\citenamefont {Schatz}\ \emph {et~al.}(2014)\citenamefont {Schatz},
  \citenamefont {Gupta}, \citenamefont {M\"oller}, \citenamefont {Beard},
  \citenamefont {Brown}, \citenamefont {Deibel}, \citenamefont {Gasques},
  \citenamefont {Hix}, \citenamefont {Keek}, \citenamefont {Lau}, \citenamefont
  {Steiner},\ and\ \citenamefont {Wiescher}}]{Nature2014}%
  \BibitemOpen
  \bibfield  {author} {\bibinfo {author} {\bibfnamefont {H.}~\bibnamefont
  {Schatz}}, \bibinfo {author} {\bibfnamefont {S.}~\bibnamefont {Gupta}},
  \bibinfo {author} {\bibfnamefont {P.}~\bibnamefont {M\"oller}}, \bibinfo
  {author} {\bibfnamefont {M.}~\bibnamefont {Beard}}, \bibinfo {author}
  {\bibfnamefont {E.~F.}\ \bibnamefont {Brown}}, \bibinfo {author}
  {\bibfnamefont {A.~T.}\ \bibnamefont {Deibel}}, \bibinfo {author}
  {\bibfnamefont {L.~R.}\ \bibnamefont {Gasques}}, \bibinfo {author}
  {\bibfnamefont {W.~R.}\ \bibnamefont {Hix}}, \bibinfo {author} {\bibfnamefont
  {L.}~\bibnamefont {Keek}}, \bibinfo {author} {\bibfnamefont {R.}~\bibnamefont
  {Lau}}, \bibinfo {author} {\bibfnamefont {A.~W.}\ \bibnamefont {Steiner}},\
  and\ \bibinfo {author} {\bibfnamefont {M.}~\bibnamefont {Wiescher}},\
  }\bibfield  {title} {\bibinfo {title} {Strong neutrino cooling by cycles of
  electron capture and beta decay in neutron star crusts},\ }\href@noop {}
  {\bibfield  {journal} {\bibinfo  {journal} {Nature}\ }\textbf {\bibinfo
  {volume} {505}},\ \bibinfo {pages} {62} (\bibinfo {year} {2014})}\BibitemShut
  {NoStop}%
\bibitem{Our_Urca_arXiv} L.-J. Wang, L. Tan, Z. Li, G. W. Misch, and Y. Sun, Phys. Rev. Lett. \textbf{127}, 172702 (2021).
\bibitem [{\citenamefont {Iwamoto}\ \emph {et~al.}(1999)\citenamefont
  {Iwamoto}, \citenamefont {Brachwitz}, \citenamefont {Nomoto}, \citenamefont
  {Kishimoto}, \citenamefont {Umeda}, \citenamefont {Hix},\ and\ \citenamefont
  {Thielemann}}]{Iwamoto_1999}%
  \BibitemOpen
  \bibfield  {author} {\bibinfo {author} {\bibfnamefont {K.}~\bibnamefont
  {Iwamoto}}, \bibinfo {author} {\bibfnamefont {F.}~\bibnamefont {Brachwitz}},
  \bibinfo {author} {\bibfnamefont {K.}~\bibnamefont {Nomoto}}, \bibinfo
  {author} {\bibfnamefont {N.}~\bibnamefont {Kishimoto}}, \bibinfo {author}
  {\bibfnamefont {H.}~\bibnamefont {Umeda}}, \bibinfo {author} {\bibfnamefont
  {W.~R.}\ \bibnamefont {Hix}},\ and\ \bibinfo {author} {\bibfnamefont {F.-K.}\
  \bibnamefont {Thielemann}},\ }\bibfield  {title} {\bibinfo {title}
  {Nucleosynthesis in chandrasekhar mass models for type ia supernovae and
  constraints on progenitor systems and burning-front propagation},\
  }\href@noop {} {\bibfield  {journal} {\bibinfo  {journal} {Astrophys. J.
  Suppl.}\ }\textbf {\bibinfo {volume} {125}},\ \bibinfo {pages} {439}
  (\bibinfo {year} {1999})}\BibitemShut {NoStop}%
\bibitem [{\citenamefont {Zegers}\ \emph {et~al.}(2008)\citenamefont {Zegers},
  \citenamefont {Brown}, \citenamefont {Akimune}, \citenamefont {Austin},
  \citenamefont {Berg}, \citenamefont {Brown}, \citenamefont {Chamulak},
  \citenamefont {Fujita}, \citenamefont {Fujiwara}, \citenamefont {Gal\`es},
  \citenamefont {Harakeh}, \citenamefont {Hashimoto}, \citenamefont {Hayami},
  \citenamefont {Hitt}, \citenamefont {Itoh}, \citenamefont {Kawabata},
  \citenamefont {Kawase}, \citenamefont {Kinoshita}, \citenamefont {Nakanishi},
  \citenamefont {Nakayama}, \citenamefont {Okumura}, \citenamefont {Shimbara},
  \citenamefont {Uchida}, \citenamefont {Ueno}, \citenamefont {Yamagata},\ and\
  \citenamefont {Yosoi}}]{Zegers2008prc}%
  \BibitemOpen
  \bibfield  {author} {\bibinfo {author} {\bibfnamefont {R.~G.~T.}\
  \bibnamefont {Zegers}}, \bibinfo {author} {\bibfnamefont {E.~F.}\
  \bibnamefont {Brown}}, \bibinfo {author} {\bibfnamefont {H.}~\bibnamefont
  {Akimune}}, \bibinfo {author} {\bibfnamefont {S.~M.}\ \bibnamefont {Austin}},
  \bibinfo {author} {\bibfnamefont {A.~M. v.~d.}\ \bibnamefont {Berg}},
  \bibinfo {author} {\bibfnamefont {B.~A.}\ \bibnamefont {Brown}}, \bibinfo
  {author} {\bibfnamefont {D.~A.}\ \bibnamefont {Chamulak}}, \bibinfo {author}
  {\bibfnamefont {Y.}~\bibnamefont {Fujita}}, \bibinfo {author} {\bibfnamefont
  {M.}~\bibnamefont {Fujiwara}}, \bibinfo {author} {\bibfnamefont
  {S.}~\bibnamefont {Gal\`es}}, \bibinfo {author} {\bibfnamefont {M.~N.}\
  \bibnamefont {Harakeh}}, \bibinfo {author} {\bibfnamefont {H.}~\bibnamefont
  {Hashimoto}}, \bibinfo {author} {\bibfnamefont {R.}~\bibnamefont {Hayami}},
  \bibinfo {author} {\bibfnamefont {G.~W.}\ \bibnamefont {Hitt}}, \bibinfo
  {author} {\bibfnamefont {M.}~\bibnamefont {Itoh}}, \bibinfo {author}
  {\bibfnamefont {T.}~\bibnamefont {Kawabata}}, \bibinfo {author}
  {\bibfnamefont {K.}~\bibnamefont {Kawase}}, \bibinfo {author} {\bibfnamefont
  {M.}~\bibnamefont {Kinoshita}}, \bibinfo {author} {\bibfnamefont
  {K.}~\bibnamefont {Nakanishi}}, \bibinfo {author} {\bibfnamefont
  {S.}~\bibnamefont {Nakayama}}, \bibinfo {author} {\bibfnamefont
  {S.}~\bibnamefont {Okumura}}, \bibinfo {author} {\bibfnamefont
  {Y.}~\bibnamefont {Shimbara}}, \bibinfo {author} {\bibfnamefont
  {M.}~\bibnamefont {Uchida}}, \bibinfo {author} {\bibfnamefont
  {H.}~\bibnamefont {Ueno}}, \bibinfo {author} {\bibfnamefont {T.}~\bibnamefont
  {Yamagata}},\ and\ \bibinfo {author} {\bibfnamefont {M.}~\bibnamefont
  {Yosoi}},\ }\bibfield  {title} {\bibinfo {title} {Gamow-teller strength for
  the analog transitions to the first
  $t=1/2,{J}^{\ensuremath{\pi}}=3/{2}^{\ensuremath{-}}$ states in
  $^{13}\mathrm{C}$ and $^{13}\mathrm{N}$ and the implications for type ia
  supernovae},\ }\href@noop {} {\bibfield  {journal} {\bibinfo  {journal}
  {Phys. Rev. C}\ }\textbf {\bibinfo {volume} {77}},\ \bibinfo {pages} {024307}
  (\bibinfo {year} {2008})}\BibitemShut {NoStop}%
\bibitem [{\citenamefont {Fujita}\ \emph {et~al.}(2011)\citenamefont {Fujita},
  \citenamefont {Rubio},\ and\ \citenamefont {Gelletly}}]{Fujita-2011-ppnp}%
  \BibitemOpen
  \bibfield  {author} {\bibinfo {author} {\bibfnamefont {Y.}~\bibnamefont
  {Fujita}}, \bibinfo {author} {\bibfnamefont {B.}~\bibnamefont {Rubio}},\ and\
  \bibinfo {author} {\bibfnamefont {W.}~\bibnamefont {Gelletly}},\ }\bibfield
  {title} {\bibinfo {title} {Spin--isospin excitations probed by strong, weak
  and electro-magnetic interactions},\ }\href@noop {} {\bibfield  {journal}
  {\bibinfo  {journal} {Prog. Part. Nucl. Phys.}\ }\textbf {\bibinfo {volume}
  {66}},\ \bibinfo {pages} {549} (\bibinfo {year} {2011})}\BibitemShut
  {NoStop}%
\bibitem [{\citenamefont {Gao}\ \emph {et~al.}(2021)\citenamefont {Gao},
  \citenamefont {Giraud}, \citenamefont {Li}, \citenamefont {Sieverding},
  \citenamefont {Zegers}, \citenamefont {Tang}, \citenamefont {Ash},
  \citenamefont {Ayyad-Limonge}, \citenamefont {Bazin}, \citenamefont {Biswas},
  \citenamefont {Brown}, \citenamefont {Chen}, \citenamefont {DeNudt},
  \citenamefont {Farris}, \citenamefont {Gabler}, \citenamefont {Gade},
  \citenamefont {Ginter}, \citenamefont {Grinder}, \citenamefont {Heger},
  \citenamefont {Hultquist}, \citenamefont {Hill}, \citenamefont {Iwasaki},
  \citenamefont {Kwan}, \citenamefont {Li}, \citenamefont {Longfellow},
  \citenamefont {Maher}, \citenamefont {Ndayisabye}, \citenamefont {Noji},
  \citenamefont {Pereira}, \citenamefont {Qi}, \citenamefont {Rebenstock},
  \citenamefont {Revel}, \citenamefont {Rhodes}, \citenamefont {Sanchez},
  \citenamefont {Schmitt}, \citenamefont {Sumithrarachchi}, \citenamefont
  {Sun},\ and\ \citenamefont {Weisshaar}}]{BSGao_2021_PRL}%
  \BibitemOpen
  \bibfield  {author} {\bibinfo {author} {\bibfnamefont {B.}~\bibnamefont
  {Gao}}, \bibinfo {author} {\bibfnamefont {S.}~\bibnamefont {Giraud}},
  \bibinfo {author} {\bibfnamefont {K.~A.}\ \bibnamefont {Li}}, \bibinfo
  {author} {\bibfnamefont {A.}~\bibnamefont {Sieverding}}, \bibinfo {author}
  {\bibfnamefont {R.~G.~T.}\ \bibnamefont {Zegers}}, \bibinfo {author}
  {\bibfnamefont {X.}~\bibnamefont {Tang}}, \bibinfo {author} {\bibfnamefont
  {J.}~\bibnamefont {Ash}}, \bibinfo {author} {\bibfnamefont {Y.}~\bibnamefont
  {Ayyad-Limonge}}, \bibinfo {author} {\bibfnamefont {D.}~\bibnamefont
  {Bazin}}, \bibinfo {author} {\bibfnamefont {S.}~\bibnamefont {Biswas}},
  \bibinfo {author} {\bibfnamefont {B.~A.}\ \bibnamefont {Brown}}, \bibinfo
  {author} {\bibfnamefont {J.}~\bibnamefont {Chen}}, \bibinfo {author}
  {\bibfnamefont {M.}~\bibnamefont {DeNudt}}, \bibinfo {author} {\bibfnamefont
  {P.}~\bibnamefont {Farris}}, \bibinfo {author} {\bibfnamefont {J.~M.}\
  \bibnamefont {Gabler}}, \bibinfo {author} {\bibfnamefont {A.}~\bibnamefont
  {Gade}}, \bibinfo {author} {\bibfnamefont {T.}~\bibnamefont {Ginter}},
  \bibinfo {author} {\bibfnamefont {M.}~\bibnamefont {Grinder}}, \bibinfo
  {author} {\bibfnamefont {A.}~\bibnamefont {Heger}}, \bibinfo {author}
  {\bibfnamefont {C.}~\bibnamefont {Hultquist}}, \bibinfo {author}
  {\bibfnamefont {A.~M.}\ \bibnamefont {Hill}}, \bibinfo {author}
  {\bibfnamefont {H.}~\bibnamefont {Iwasaki}}, \bibinfo {author} {\bibfnamefont
  {E.}~\bibnamefont {Kwan}}, \bibinfo {author} {\bibfnamefont {J.}~\bibnamefont
  {Li}}, \bibinfo {author} {\bibfnamefont {B.}~\bibnamefont {Longfellow}},
  \bibinfo {author} {\bibfnamefont {C.}~\bibnamefont {Maher}}, \bibinfo
  {author} {\bibfnamefont {F.}~\bibnamefont {Ndayisabye}}, \bibinfo {author}
  {\bibfnamefont {S.}~\bibnamefont {Noji}}, \bibinfo {author} {\bibfnamefont
  {J.}~\bibnamefont {Pereira}}, \bibinfo {author} {\bibfnamefont
  {C.}~\bibnamefont {Qi}}, \bibinfo {author} {\bibfnamefont {J.}~\bibnamefont
  {Rebenstock}}, \bibinfo {author} {\bibfnamefont {A.}~\bibnamefont {Revel}},
  \bibinfo {author} {\bibfnamefont {D.}~\bibnamefont {Rhodes}}, \bibinfo
  {author} {\bibfnamefont {A.}~\bibnamefont {Sanchez}}, \bibinfo {author}
  {\bibfnamefont {J.}~\bibnamefont {Schmitt}}, \bibinfo {author} {\bibfnamefont
  {C.}~\bibnamefont {Sumithrarachchi}}, \bibinfo {author} {\bibfnamefont
  {B.~H.}\ \bibnamefont {Sun}},\ and\ \bibinfo {author} {\bibfnamefont
  {D.}~\bibnamefont {Weisshaar}},\ }\bibfield  {title} {\bibinfo {title} {New
  $^{59}\mathrm{Fe}$ stellar decay rate with implications for the
  $^{60}\mathrm{Fe}$ radioactivity in massive stars},\ }\href
  {https://doi.org/10.1103/PhysRevLett.126.152701} {\bibfield  {journal}
  {\bibinfo  {journal} {Phys. Rev. Lett.}\ }\textbf {\bibinfo {volume} {126}},\
  \bibinfo {pages} {152701} (\bibinfo {year} {2021})}\BibitemShut {NoStop}%
\bibitem [{\citenamefont {Tan}\ \emph {et~al.}(2020)\citenamefont {Tan},
  \citenamefont {Liu}, \citenamefont {Wang}, \citenamefont {Li},\ and\
  \citenamefont {Sun}}]{Tan_2020_PLB}%
  \BibitemOpen
  \bibfield  {author} {\bibinfo {author} {\bibfnamefont {L.}~\bibnamefont
  {Tan}}, \bibinfo {author} {\bibfnamefont {Y.-X.}\ \bibnamefont {Liu}},
  \bibinfo {author} {\bibfnamefont {L.-J.}\ \bibnamefont {Wang}}, \bibinfo
  {author} {\bibfnamefont {Z.}~\bibnamefont {Li}},\ and\ \bibinfo {author}
  {\bibfnamefont {Y.}~\bibnamefont {Sun}},\ }\bibfield  {title} {\bibinfo
  {title} {A novel method for stellar electron-capture rates of excited nuclear
  states},\ }\href
  {https://doi.org/https://doi.org/10.1016/j.physletb.2020.135432} {\bibfield
  {journal} {\bibinfo  {journal} {Phys. Lett. B}\ }\textbf {\bibinfo {volume}
  {805}},\ \bibinfo {pages} {135432} (\bibinfo {year} {2020})}\BibitemShut
  {NoStop}%
\bibitem [{\citenamefont {Oda}\ \emph {et~al.}(1994)\citenamefont {Oda},
  \citenamefont {Hino}, \citenamefont {Muto}, \citenamefont {Takahara},\ and\
  \citenamefont {Sato}}]{Oda-1994}%
  \BibitemOpen
  \bibfield  {author} {\bibinfo {author} {\bibfnamefont {T.}~\bibnamefont
  {Oda}}, \bibinfo {author} {\bibfnamefont {M.}~\bibnamefont {Hino}}, \bibinfo
  {author} {\bibfnamefont {K.}~\bibnamefont {Muto}}, \bibinfo {author}
  {\bibfnamefont {M.}~\bibnamefont {Takahara}},\ and\ \bibinfo {author}
  {\bibfnamefont {K.}~\bibnamefont {Sato}},\ }\bibfield  {title} {\bibinfo
  {title} {Rate tables for the weak processes of $sd$-shell nuclei in stellar
  matter},\ }\href@noop {} {\bibfield  {journal} {\bibinfo  {journal} {At. Data
  Nucl. Data Tables}\ }\textbf {\bibinfo {volume} {56}},\ \bibinfo {pages}
  {231} (\bibinfo {year} {1994})}\BibitemShut {NoStop}%
\bibitem [{\citenamefont {Mart\'{\i}nez-Pinedo}\ \emph
  {et~al.}(2014)\citenamefont {Mart\'{\i}nez-Pinedo}, \citenamefont {Lam},
  \citenamefont {Langanke}, \citenamefont {Zegers},\ and\ \citenamefont
  {Sullivan}}]{Gabriel-2014}%
  \BibitemOpen
  \bibfield  {author} {\bibinfo {author} {\bibfnamefont {G.}~\bibnamefont
  {Mart\'{\i}nez-Pinedo}}, \bibinfo {author} {\bibfnamefont {Y.~H.}\
  \bibnamefont {Lam}}, \bibinfo {author} {\bibfnamefont {K.}~\bibnamefont
  {Langanke}}, \bibinfo {author} {\bibfnamefont {R.~G.~T.}\ \bibnamefont
  {Zegers}},\ and\ \bibinfo {author} {\bibfnamefont {C.}~\bibnamefont
  {Sullivan}},\ }\bibfield  {title} {\bibinfo {title} {Astrophysical
  weak-interaction rates for selected $a=20$ and $a=24$ nuclei},\ }\href@noop
  {} {\bibfield  {journal} {\bibinfo  {journal} {Phys. Rev. C}\ }\textbf
  {\bibinfo {volume} {89}},\ \bibinfo {pages} {045806} (\bibinfo {year}
  {2014})}\BibitemShut {NoStop}%
\bibitem [{\citenamefont {Misch}\ \emph {et~al.}(2014)\citenamefont {Misch},
  \citenamefont {Fuller},\ and\ \citenamefont {Brown}}]{wendell-2014}%
  \BibitemOpen
  \bibfield  {author} {\bibinfo {author} {\bibfnamefont {G.~W.}\ \bibnamefont
  {Misch}}, \bibinfo {author} {\bibfnamefont {G.~M.}\ \bibnamefont {Fuller}},\
  and\ \bibinfo {author} {\bibfnamefont {B.~A.}\ \bibnamefont {Brown}},\
  }\bibfield  {title} {\bibinfo {title} {Modification of the brink-axel
  hypothesis for high-temperature nuclear weak interactions},\ }\href@noop {}
  {\bibfield  {journal} {\bibinfo  {journal} {Phys. Rev. C}\ }\textbf {\bibinfo
  {volume} {90}},\ \bibinfo {pages} {065808} (\bibinfo {year}
  {2014})}\BibitemShut {NoStop}%
\bibitem [{\citenamefont {Caurier}\ \emph {et~al.}(1999)\citenamefont
  {Caurier}, \citenamefont {Langanke}, \citenamefont {Mart\'{\i}nez-Pinedo},\
  and\ \citenamefont {Nowacki}}]{Caurier-1999}%
  \BibitemOpen
  \bibfield  {author} {\bibinfo {author} {\bibfnamefont {E.}~\bibnamefont
  {Caurier}}, \bibinfo {author} {\bibfnamefont {K.}~\bibnamefont {Langanke}},
  \bibinfo {author} {\bibfnamefont {G.}~\bibnamefont {Mart\'{\i}nez-Pinedo}},\
  and\ \bibinfo {author} {\bibfnamefont {F.}~\bibnamefont {Nowacki}},\
  }\bibfield  {title} {\bibinfo {title} {Shell-model calculations of stellar
  weak interaction rates. i. gamow-teller distributions and spectra of nuclei
  in the mass range a = 45--65},\ }\href@noop {} {\bibfield  {journal}
  {\bibinfo  {journal} {Nucl. Phys. A}\ }\textbf {\bibinfo {volume} {653}},\
  \bibinfo {pages} {439} (\bibinfo {year} {1999})}\BibitemShut {NoStop}%
\bibitem [{\citenamefont {Langanke}\ and\ \citenamefont
  {Mart\'{\i}nez-Pinedo}(2000)}]{Langanke2000}%
  \BibitemOpen
  \bibfield  {author} {\bibinfo {author} {\bibfnamefont {K.}~\bibnamefont
  {Langanke}}\ and\ \bibinfo {author} {\bibfnamefont {G.}~\bibnamefont
  {Mart\'{\i}nez-Pinedo}},\ }\bibfield  {title} {\bibinfo {title} {Shell-model
  calculations of stellar weak interaction rates: Ii.},\ }\href@noop {}
  {\bibfield  {journal} {\bibinfo  {journal} {Nucl. Phys. A}\ }\textbf
  {\bibinfo {volume} {673}},\ \bibinfo {pages} {481} (\bibinfo {year}
  {2000})}\BibitemShut {NoStop}%
\bibitem [{\citenamefont {Langanke}\ and\ \citenamefont
  {Mart\'{\i}nez-Pinedo}(2001)}]{Langanke2001}%
  \BibitemOpen
  \bibfield  {author} {\bibinfo {author} {\bibfnamefont {K.}~\bibnamefont
  {Langanke}}\ and\ \bibinfo {author} {\bibfnamefont {G.}~\bibnamefont
  {Mart\'{\i}nez-Pinedo}},\ }\bibfield  {title} {\bibinfo {title} {Rates tables
  for the weak processes of pf-shell nuclei in stellar environments},\
  }\href@noop {} {\bibfield  {journal} {\bibinfo  {journal} {At. Data Nucl.
  Data Tables}\ }\textbf {\bibinfo {volume} {79}},\ \bibinfo {pages} {1}
  (\bibinfo {year} {2001})}\BibitemShut {NoStop}%
\bibitem [{\citenamefont {Langanke}\ \emph {et~al.}(2001)\citenamefont
  {Langanke}, \citenamefont {Kolbe},\ and\ \citenamefont
  {Dean}}]{Langanke_2001_PRC}%
  \BibitemOpen
  \bibfield  {author} {\bibinfo {author} {\bibfnamefont {K.}~\bibnamefont
  {Langanke}}, \bibinfo {author} {\bibfnamefont {E.}~\bibnamefont {Kolbe}},\
  and\ \bibinfo {author} {\bibfnamefont {D.~J.}\ \bibnamefont {Dean}},\
  }\bibfield  {title} {\bibinfo {title} {Unblocking of the gamow-teller
  strength in stellar electron capture on neutron-rich germanium isotopes},\
  }\href {https://doi.org/10.1103/PhysRevC.63.032801} {\bibfield  {journal}
  {\bibinfo  {journal} {Phys. Rev. C}\ }\textbf {\bibinfo {volume} {63}},\
  \bibinfo {pages} {032801(R)} (\bibinfo {year} {2001})}\BibitemShut {NoStop}%
\bibitem [{\citenamefont {Langanke}\ \emph {et~al.}(2003)\citenamefont
  {Langanke}, \citenamefont {Mart\'{\i}nez-Pinedo}, \citenamefont {Sampaio},
  \citenamefont {Dean}, \citenamefont {Hix}, \citenamefont {Messer},
  \citenamefont {Mezzacappa}, \citenamefont {Liebend\"orfer}, \citenamefont
  {Janka},\ and\ \citenamefont {Rampp}}]{Langanke_2003_PRL}%
  \BibitemOpen
  \bibfield  {author} {\bibinfo {author} {\bibfnamefont {K.}~\bibnamefont
  {Langanke}}, \bibinfo {author} {\bibfnamefont {G.}~\bibnamefont
  {Mart\'{\i}nez-Pinedo}}, \bibinfo {author} {\bibfnamefont {J.~M.}\
  \bibnamefont {Sampaio}}, \bibinfo {author} {\bibfnamefont {D.~J.}\
  \bibnamefont {Dean}}, \bibinfo {author} {\bibfnamefont {W.~R.}\ \bibnamefont
  {Hix}}, \bibinfo {author} {\bibfnamefont {O.~E.~B.}\ \bibnamefont {Messer}},
  \bibinfo {author} {\bibfnamefont {A.}~\bibnamefont {Mezzacappa}}, \bibinfo
  {author} {\bibfnamefont {M.}~\bibnamefont {Liebend\"orfer}}, \bibinfo
  {author} {\bibfnamefont {H.-T.}\ \bibnamefont {Janka}},\ and\ \bibinfo
  {author} {\bibfnamefont {M.}~\bibnamefont {Rampp}},\ }\bibfield  {title}
  {\bibinfo {title} {Electron capture rates on nuclei and implications for
  stellar core collapse},\ }\href
  {https://doi.org/10.1103/PhysRevLett.90.241102} {\bibfield  {journal}
  {\bibinfo  {journal} {Phys. Rev. Lett.}\ }\textbf {\bibinfo {volume} {90}},\
  \bibinfo {pages} {241102} (\bibinfo {year} {2003})}\BibitemShut {NoStop}%
\bibitem [{\citenamefont {Engel}\ \emph {et~al.}(1999)\citenamefont {Engel},
  \citenamefont {Bender}, \citenamefont {Dobaczewski}, \citenamefont
  {Nazarewicz},\ and\ \citenamefont {Surman}}]{jon-1999}%
  \BibitemOpen
  \bibfield  {author} {\bibinfo {author} {\bibfnamefont {J.}~\bibnamefont
  {Engel}}, \bibinfo {author} {\bibfnamefont {M.}~\bibnamefont {Bender}},
  \bibinfo {author} {\bibfnamefont {J.}~\bibnamefont {Dobaczewski}}, \bibinfo
  {author} {\bibfnamefont {W.}~\bibnamefont {Nazarewicz}},\ and\ \bibinfo
  {author} {\bibfnamefont {R.}~\bibnamefont {Surman}},\ }\bibfield  {title}
  {\bibinfo {title} {$\beta$ decay rates of r-process waiting-point nuclei in a
  self-consistent approach},\ }\href@noop {} {\bibfield  {journal} {\bibinfo
  {journal} {Phys. Rev. C}\ }\textbf {\bibinfo {volume} {60}},\ \bibinfo
  {pages} {014302} (\bibinfo {year} {1999})}\BibitemShut {NoStop}%
\bibitem [{\citenamefont {Paar}\ \emph {et~al.}(2009)\citenamefont {Paar},
  \citenamefont {Col\`o}, \citenamefont {Khan},\ and\ \citenamefont
  {Vretenar}}]{ec-Paar}%
  \BibitemOpen
  \bibfield  {author} {\bibinfo {author} {\bibfnamefont {N.}~\bibnamefont
  {Paar}}, \bibinfo {author} {\bibfnamefont {G.}~\bibnamefont {Col\`o}},
  \bibinfo {author} {\bibfnamefont {E.}~\bibnamefont {Khan}},\ and\ \bibinfo
  {author} {\bibfnamefont {D.}~\bibnamefont {Vretenar}},\ }\bibfield  {title}
  {\bibinfo {title} {Calculation of stellar electron-capture cross sections on
  nuclei based on microscopic skyrme functionals},\ }\href@noop {} {\bibfield
  {journal} {\bibinfo  {journal} {Phys. Rev. C}\ }\textbf {\bibinfo {volume}
  {80}},\ \bibinfo {pages} {055801} (\bibinfo {year} {2009})}\BibitemShut
  {NoStop}%
\bibitem [{\citenamefont {Bai}\ \emph {et~al.}(2010)\citenamefont {Bai},
  \citenamefont {Zhang}, \citenamefont {Sagawa}, \citenamefont {Zhang},
  \citenamefont {Col\`o},\ and\ \citenamefont {Xu}}]{Bai-2010}%
  \BibitemOpen
  \bibfield  {author} {\bibinfo {author} {\bibfnamefont {C.~L.}\ \bibnamefont
  {Bai}}, \bibinfo {author} {\bibfnamefont {H.~Q.}\ \bibnamefont {Zhang}},
  \bibinfo {author} {\bibfnamefont {H.}~\bibnamefont {Sagawa}}, \bibinfo
  {author} {\bibfnamefont {X.~Z.}\ \bibnamefont {Zhang}}, \bibinfo {author}
  {\bibfnamefont {G.}~\bibnamefont {Col\`o}},\ and\ \bibinfo {author}
  {\bibfnamefont {F.~R.}\ \bibnamefont {Xu}},\ }\bibfield  {title} {\bibinfo
  {title} {Effect of the tensor force on the charge exchange spin-dipole
  excitations of $^{208}\mathrm{Pb}$},\ }\href@noop {} {\bibfield  {journal}
  {\bibinfo  {journal} {Phys. Rev. Lett.}\ }\textbf {\bibinfo {volume} {105}},\
  \bibinfo {pages} {072501} (\bibinfo {year} {2010})}\BibitemShut {NoStop}%
\bibitem [{\citenamefont {Dzhioev}\ \emph {et~al.}(2010)\citenamefont
  {Dzhioev}, \citenamefont {Vdovin}, \citenamefont {Ponomarev}, \citenamefont
  {Wambach}, \citenamefont {Langanke},\ and\ \citenamefont
  {Mart\'{\i}nez-Pinedo}}]{Dzhioev2010}%
  \BibitemOpen
  \bibfield  {author} {\bibinfo {author} {\bibfnamefont {A.~A.}\ \bibnamefont
  {Dzhioev}}, \bibinfo {author} {\bibfnamefont {A.~I.}\ \bibnamefont {Vdovin}},
  \bibinfo {author} {\bibfnamefont {V.~Y.}\ \bibnamefont {Ponomarev}}, \bibinfo
  {author} {\bibfnamefont {J.}~\bibnamefont {Wambach}}, \bibinfo {author}
  {\bibfnamefont {K.}~\bibnamefont {Langanke}},\ and\ \bibinfo {author}
  {\bibfnamefont {G.}~\bibnamefont {Mart\'{\i}nez-Pinedo}},\ }\bibfield
  {title} {\bibinfo {title} {Gamow-teller strength distributions at finite
  temperatures and electron capture in stellar environments},\ }\href@noop {}
  {\bibfield  {journal} {\bibinfo  {journal} {Phys. Rev. C}\ }\textbf {\bibinfo
  {volume} {81}},\ \bibinfo {pages} {015804} (\bibinfo {year}
  {2010})}\BibitemShut {NoStop}%
\bibitem [{\citenamefont {Niu}\ \emph {et~al.}(2011)\citenamefont {Niu},
  \citenamefont {Paar}, \citenamefont {Vretenar},\ and\ \citenamefont
  {Meng}}]{ec-Niu}%
  \BibitemOpen
  \bibfield  {author} {\bibinfo {author} {\bibfnamefont {Y.~F.}\ \bibnamefont
  {Niu}}, \bibinfo {author} {\bibfnamefont {N.}~\bibnamefont {Paar}}, \bibinfo
  {author} {\bibfnamefont {D.}~\bibnamefont {Vretenar}},\ and\ \bibinfo
  {author} {\bibfnamefont {J.}~\bibnamefont {Meng}},\ }\bibfield  {title}
  {\bibinfo {title} {Stellar electron-capture rates calculated with the
  finite-temperature relativistic random-phase approximation},\ }\href@noop {}
  {\bibfield  {journal} {\bibinfo  {journal} {Phys. Rev. C}\ }\textbf {\bibinfo
  {volume} {83}},\ \bibinfo {pages} {045807} (\bibinfo {year}
  {2011})}\BibitemShut {NoStop}%
\bibitem [{\citenamefont {Fang}\ \emph {et~al.}(2013)\citenamefont {Fang},
  \citenamefont {Brown},\ and\ \citenamefont {Suzuki}}]{fang-2013}%
  \BibitemOpen
  \bibfield  {author} {\bibinfo {author} {\bibfnamefont {D.-L.}\ \bibnamefont
  {Fang}}, \bibinfo {author} {\bibfnamefont {B.~A.}\ \bibnamefont {Brown}},\
  and\ \bibinfo {author} {\bibfnamefont {T.}~\bibnamefont {Suzuki}},\
  }\bibfield  {title} {\bibinfo {title} {$\beta$ decay properties for
  neutron-rich kr-tc isotopes from deformed pn-quasiparticle random-phase
  approximation calculations with realistic forces},\ }\href@noop {} {\bibfield
   {journal} {\bibinfo  {journal} {Phys. Rev. C}\ }\textbf {\bibinfo {volume}
  {88}},\ \bibinfo {pages} {024314} (\bibinfo {year} {2013})}\BibitemShut
  {NoStop}%
\bibitem [{\citenamefont {Niu}\ \emph {et~al.}(2013)\citenamefont {Niu},
  \citenamefont {Niu}, \citenamefont {Liang}, \citenamefont {Long},
  \citenamefont {Nik\v{s}i\'{c}}, \citenamefont {Vretenar},\ and\ \citenamefont
  {Meng}}]{Niu-2013}%
  \BibitemOpen
  \bibfield  {author} {\bibinfo {author} {\bibfnamefont {Z.~M.}\ \bibnamefont
  {Niu}}, \bibinfo {author} {\bibfnamefont {Y.~F.}\ \bibnamefont {Niu}},
  \bibinfo {author} {\bibfnamefont {H.~Z.}\ \bibnamefont {Liang}}, \bibinfo
  {author} {\bibfnamefont {W.~H.}\ \bibnamefont {Long}}, \bibinfo {author}
  {\bibnamefont {Nik\v{s}i\'{c}}}, \bibinfo {author} {\bibfnamefont
  {D.}~\bibnamefont {Vretenar}},\ and\ \bibinfo {author} {\bibfnamefont
  {J.}~\bibnamefont {Meng}},\ }\bibfield  {title} {\bibinfo {title}
  {$\beta$-decay half-lives of neutron-rich nuclei and matter flow in the
  r-process},\ }\href@noop {} {\bibfield  {journal} {\bibinfo  {journal} {Phys.
  Lett. B}\ }\textbf {\bibinfo {volume} {723}},\ \bibinfo {pages} {172}
  (\bibinfo {year} {2013})}\BibitemShut {NoStop}%
\bibitem [{\citenamefont {Sarriguren}(2013)}]{Sarriguren2013}%
  \BibitemOpen
  \bibfield  {author} {\bibinfo {author} {\bibfnamefont {P.}~\bibnamefont
  {Sarriguren}},\ }\bibfield  {title} {\bibinfo {title} {Stellar
  electron-capture rates in $pf$-shell nuclei from quasiparticle-random-phase
  approximatoin calculations},\ }\href@noop {} {\bibfield  {journal} {\bibinfo
  {journal} {Phys. Rev. C}\ }\textbf {\bibinfo {volume} {87}},\ \bibinfo
  {pages} {045801} (\bibinfo {year} {2013})}\BibitemShut {NoStop}%
\bibitem [{\citenamefont {Robin}\ and\ \citenamefont
  {Litvinova}(2019)}]{Robin-2019}%
  \BibitemOpen
  \bibfield  {author} {\bibinfo {author} {\bibfnamefont {C.}~\bibnamefont
  {Robin}}\ and\ \bibinfo {author} {\bibfnamefont {E.}~\bibnamefont
  {Litvinova}},\ }\bibfield  {title} {\bibinfo {title} {Time-reversed
  particle-vibration loops and nuclear gamow-teller response},\ }\href@noop {}
  {\bibfield  {journal} {\bibinfo  {journal} {Phys. Rev. Lett.}\ }\textbf
  {\bibinfo {volume} {123}},\ \bibinfo {pages} {202501} (\bibinfo {year}
  {2019})}\BibitemShut {NoStop}%
\bibitem [{\citenamefont {Ejiri}\ \emph {et~al.}(2019)\citenamefont {Ejiri},
  \citenamefont {Suhonen},\ and\ \citenamefont {Zuber}}]{Suhonen-2019-review}%
  \BibitemOpen
  \bibfield  {author} {\bibinfo {author} {\bibfnamefont {H.}~\bibnamefont
  {Ejiri}}, \bibinfo {author} {\bibfnamefont {J.}~\bibnamefont {Suhonen}},\
  and\ \bibinfo {author} {\bibfnamefont {K.}~\bibnamefont {Zuber}},\ }\bibfield
   {title} {\bibinfo {title} {Neutrino--nuclear responses for astro-neutrinos,
  single beta decays and double beta decays},\ }\href@noop {} {\bibfield
  {journal} {\bibinfo  {journal} {Phys. Rep.}\ }\textbf {\bibinfo {volume}
  {797}},\ \bibinfo {pages} {1} (\bibinfo {year} {2019})}\BibitemShut {NoStop}%
\bibitem [{\citenamefont {Brink}(1955)}]{Brink1955}%
  \BibitemOpen
  \bibfield  {author} {\bibinfo {author} {\bibfnamefont {D.~M.}\ \bibnamefont
  {Brink}},\ }\href@noop {} {\bibfield  {journal} {\bibinfo  {journal} {Ph.D.
  thesis, Oxford University}\ } (\bibinfo {year} {1955})}\BibitemShut {NoStop}%
\bibitem [{\citenamefont {Axel}(1962)}]{Axel1962}%
  \BibitemOpen
  \bibfield  {author} {\bibinfo {author} {\bibfnamefont {P.}~\bibnamefont
  {Axel}},\ }\bibfield  {title} {\bibinfo {title} {Electric dipole ground-state
  transition width strength function and 7-mev photon interactions},\
  }\href@noop {} {\bibfield  {journal} {\bibinfo  {journal} {Phys. Rev.}\
  }\textbf {\bibinfo {volume} {126}},\ \bibinfo {pages} {671} (\bibinfo {year}
  {1962})}\BibitemShut {NoStop}%
\bibitem [{\citenamefont {Haxton}\ and\ \citenamefont
  {Henley}(1995)}]{Haxton-book}%
  \BibitemOpen
  \bibfield  {author} {\bibinfo {author} {\bibfnamefont {W.~C.}\ \bibnamefont
  {Haxton}}\ and\ \bibinfo {author} {\bibfnamefont {E.~M.}\ \bibnamefont
  {Henley}},\ }\href@noop {} {\emph {\bibinfo {title} {Symmetries and
  fundamental interactions in nuclei}}}\ (\bibinfo  {publisher} {World
  Scientific (Singapore)},\ \bibinfo {year} {1995})\BibitemShut {NoStop}%
\bibitem{new_gAgV} B. M\"arkisch, H. Mest, H. Saul, et al., Phys. Rev. Lett. \textbf{122}, 242501 (2019). 
\bibitem [{\citenamefont {Men\'endez}\ \emph {et~al.}(2011)\citenamefont
  {Men\'endez}, \citenamefont {Gazit},\ and\ \citenamefont
  {Schwenk}}]{Javier2011PRL}%
  \BibitemOpen
  \bibfield  {author} {\bibinfo {author} {\bibfnamefont {J.}~\bibnamefont
  {Men\'endez}}, \bibinfo {author} {\bibfnamefont {D.}~\bibnamefont {Gazit}},\
  and\ \bibinfo {author} {\bibfnamefont {A.}~\bibnamefont {Schwenk}},\
  }\bibfield  {title} {\bibinfo {title} {Chiral two-body currents in nuclei:
  gamow-teller transitions and neutrinoless double-beta decay},\ }\href@noop {}
  {\bibfield  {journal} {\bibinfo  {journal} {Phys. Rev. Lett.}\ }\textbf
  {\bibinfo {volume} {107}},\ \bibinfo {pages} {062501} (\bibinfo {year}
  {2011})}\BibitemShut {NoStop}%
\bibitem [{\citenamefont {Wang}\ \emph
  {et~al.}(2018{\natexlab{a}})\citenamefont {Wang}, \citenamefont {Engel},\
  and\ \citenamefont {Yao}}]{Wang-0vbb-2018}%
  \BibitemOpen
  \bibfield  {author} {\bibinfo {author} {\bibfnamefont {L.-J.}\ \bibnamefont
  {Wang}}, \bibinfo {author} {\bibfnamefont {J.}~\bibnamefont {Engel}},\ and\
  \bibinfo {author} {\bibfnamefont {J.~M.}\ \bibnamefont {Yao}},\ }\bibfield
  {title} {\bibinfo {title} {Quenching of nuclear matrix elements for
  $0\ensuremath{\nu}\ensuremath{\beta}\ensuremath{\beta}$ decay by chiral
  two-body currents},\ }\href@noop {} {\bibfield  {journal} {\bibinfo
  {journal} {Phys. Rev. C}\ }\textbf {\bibinfo {volume} {98}},\ \bibinfo
  {pages} {031301(R)} (\bibinfo {year} {2018}{\natexlab{a}})}\BibitemShut
  {NoStop}%
\bibitem [{\citenamefont {Gysbers}\ \emph {et~al.}(2019)\citenamefont
  {Gysbers}, \citenamefont {Hagen}, \citenamefont {Holt}, \citenamefont
  {Jansen}, \citenamefont {Morris}, \citenamefont {Navr\'atil}, \citenamefont
  {Papenbrock}, \citenamefont {Quaglioni}, \citenamefont {Schwenk},
  \citenamefont {Stroberg},\ and\ \citenamefont {Wendt}}]{Nature2019}%
  \BibitemOpen
  \bibfield  {author} {\bibinfo {author} {\bibfnamefont {P.}~\bibnamefont
  {Gysbers}}, \bibinfo {author} {\bibfnamefont {G.}~\bibnamefont {Hagen}},
  \bibinfo {author} {\bibfnamefont {J.~D.}\ \bibnamefont {Holt}}, \bibinfo
  {author} {\bibfnamefont {G.~R.}\ \bibnamefont {Jansen}}, \bibinfo {author}
  {\bibfnamefont {T.~D.}\ \bibnamefont {Morris}}, \bibinfo {author}
  {\bibfnamefont {P.}~\bibnamefont {Navr\'atil}}, \bibinfo {author}
  {\bibfnamefont {T.}~\bibnamefont {Papenbrock}}, \bibinfo {author}
  {\bibfnamefont {S.}~\bibnamefont {Quaglioni}}, \bibinfo {author}
  {\bibfnamefont {A.}~\bibnamefont {Schwenk}}, \bibinfo {author} {\bibfnamefont
  {S.~R.}\ \bibnamefont {Stroberg}},\ and\ \bibinfo {author} {\bibfnamefont
  {K.~A.}\ \bibnamefont {Wendt}},\ }\bibfield  {title} {\bibinfo {title}
  {Discrepancy between experimental and theoretical beta-decay rates resolved
  from first principles},\ }\href@noop {} {\bibfield  {journal} {\bibinfo
  {journal} {Nature Physics}\ }\textbf {\bibinfo {volume} {15}},\ \bibinfo
  {pages} {428} (\bibinfo {year} {2019})}\BibitemShut {NoStop}%
\bibitem [{\citenamefont {Bodenstein-Dresler}\ \emph
  {et~al.}(2020)\citenamefont {Bodenstein-Dresler}, \citenamefont {Chu},
  \citenamefont {Gehre}, \citenamefont {G{\"o}{\ss}ling}, \citenamefont
  {Heimbold}, \citenamefont {Herrmann}, \citenamefont {Hodak}, \citenamefont
  {Kostensalo}, \citenamefont {Kr{\"o}ninger}, \citenamefont {K{\"u}ttler},
  \citenamefont {Nitsch}, \citenamefont {Quante}, \citenamefont {Rukhadze},
  \citenamefont {Stekl}, \citenamefont {Suhonen}, \citenamefont {Tebr{\"u}gge},
  \citenamefont {Temminghoff}, \citenamefont {Volkmer}, \citenamefont
  {Zatschler},\ and\ \citenamefont {Zuber}}]{quenching_2020_PLB}%
  \BibitemOpen
  \bibfield  {author} {\bibinfo {author} {\bibfnamefont {L.}~\bibnamefont
  {Bodenstein-Dresler}}, \bibinfo {author} {\bibfnamefont {Y.}~\bibnamefont
  {Chu}}, \bibinfo {author} {\bibfnamefont {D.}~\bibnamefont {Gehre}}, \bibinfo
  {author} {\bibfnamefont {C.}~\bibnamefont {G{\"o}{\ss}ling}}, \bibinfo
  {author} {\bibfnamefont {A.}~\bibnamefont {Heimbold}}, \bibinfo {author}
  {\bibfnamefont {C.}~\bibnamefont {Herrmann}}, \bibinfo {author}
  {\bibfnamefont {R.}~\bibnamefont {Hodak}}, \bibinfo {author} {\bibfnamefont
  {J.}~\bibnamefont {Kostensalo}}, \bibinfo {author} {\bibfnamefont
  {K.}~\bibnamefont {Kr{\"o}ninger}}, \bibinfo {author} {\bibfnamefont
  {J.}~\bibnamefont {K{\"u}ttler}}, \bibinfo {author} {\bibfnamefont
  {C.}~\bibnamefont {Nitsch}}, \bibinfo {author} {\bibfnamefont
  {T.}~\bibnamefont {Quante}}, \bibinfo {author} {\bibfnamefont
  {E.}~\bibnamefont {Rukhadze}}, \bibinfo {author} {\bibfnamefont
  {I.}~\bibnamefont {Stekl}}, \bibinfo {author} {\bibfnamefont
  {J.}~\bibnamefont {Suhonen}}, \bibinfo {author} {\bibfnamefont
  {J.}~\bibnamefont {Tebr{\"u}gge}}, \bibinfo {author} {\bibfnamefont
  {R.}~\bibnamefont {Temminghoff}}, \bibinfo {author} {\bibfnamefont
  {J.}~\bibnamefont {Volkmer}}, \bibinfo {author} {\bibfnamefont
  {S.}~\bibnamefont {Zatschler}},\ and\ \bibinfo {author} {\bibfnamefont
  {K.}~\bibnamefont {Zuber}},\ }\bibfield  {title} {\bibinfo {title} {Quenching
  of $g_{\mathrm{a}}$ deduced from the $\beta$-spectrum shape of
  $^{113}\mathrm{Cd}$ measured with the $\mathrm{COBRA}$ experiment},\ }\href
  {https://doi.org/https://doi.org/10.1016/j.physletb.2019.135092} {\bibfield
  {journal} {\bibinfo  {journal} {Phys. Lett. B}\ }\textbf {\bibinfo {volume}
  {800}},\ \bibinfo {pages} {135092} (\bibinfo {year} {2020})}\BibitemShut
  {NoStop}%
\bibitem [{\citenamefont {Epelbaum}\ \emph {et~al.}(2009)\citenamefont
  {Epelbaum}, \citenamefont {Hammer},\ and\ \citenamefont
  {Mei{\ss}ner}}]{ChiralEFT2009}%
  \BibitemOpen
  \bibfield  {author} {\bibinfo {author} {\bibfnamefont {E.}~\bibnamefont
  {Epelbaum}}, \bibinfo {author} {\bibfnamefont {H.-W.}\ \bibnamefont
  {Hammer}},\ and\ \bibinfo {author} {\bibfnamefont {U.-G.}\ \bibnamefont
  {Mei{\ss}ner}},\ }\bibfield  {title} {\bibinfo {title} {Modern theory of
  nuclear forces},\ }\href@noop {} {\bibfield  {journal} {\bibinfo  {journal}
  {Rev. Mod. Phys.}\ }\textbf {\bibinfo {volume} {81}},\ \bibinfo {pages}
  {1773} (\bibinfo {year} {2009})}\BibitemShut {NoStop}%
\bibitem [{\citenamefont {Klos}\ \emph {et~al.}(2017)\citenamefont {Klos},
  \citenamefont {Carbone}, \citenamefont {Hebeler}, \citenamefont
  {Men{\'e}ndez},\ and\ \citenamefont {Schwenk}}]{Klos_2017_EPJA}%
  \BibitemOpen
  \bibfield  {author} {\bibinfo {author} {\bibfnamefont {P.}~\bibnamefont
  {Klos}}, \bibinfo {author} {\bibfnamefont {A.}~\bibnamefont {Carbone}},
  \bibinfo {author} {\bibfnamefont {K.}~\bibnamefont {Hebeler}}, \bibinfo
  {author} {\bibfnamefont {J.}~\bibnamefont {Men{\'e}ndez}},\ and\ \bibinfo
  {author} {\bibfnamefont {A.}~\bibnamefont {Schwenk}},\ }\bibfield  {title}
  {\bibinfo {title} {Uncertainties in constraining low-energy constants from
  $^{3}\text{H}$ $\beta$ decay},\ }\href
  {https://doi.org/10.1140/epja/i2017-12357-7} {\bibfield  {journal} {\bibinfo
  {journal} {Eur. Phys. J. A.}\ }\textbf {\bibinfo {volume} {53}},\ \bibinfo
  {pages} {168} (\bibinfo {year} {2017})}\BibitemShut {NoStop}%
\bibitem [{\citenamefont {Sun}\ and\ \citenamefont {Feng}(1996)}]{PSM-Sun}%
  \BibitemOpen
  \bibfield  {author} {\bibinfo {author} {\bibfnamefont {Y.}~\bibnamefont
  {Sun}}\ and\ \bibinfo {author} {\bibfnamefont {D.~H.}\ \bibnamefont {Feng}},\
  }\bibfield  {title} {\bibinfo {title} {High spin spectroscopy with the
  projected shell model},\ }\href@noop {} {\bibfield  {journal} {\bibinfo
  {journal} {Phys. Rep.}\ }\textbf {\bibinfo {volume} {264}},\ \bibinfo {pages}
  {375} (\bibinfo {year} {1996})}\BibitemShut {NoStop}%
\bibitem [{\citenamefont {Gao}\ \emph {et~al.}(2006)\citenamefont {Gao},
  \citenamefont {Sun},\ and\ \citenamefont {Chen}}]{Gao2006}%
  \BibitemOpen
  \bibfield  {author} {\bibinfo {author} {\bibfnamefont {Z.-C.}\ \bibnamefont
  {Gao}}, \bibinfo {author} {\bibfnamefont {Y.}~\bibnamefont {Sun}},\ and\
  \bibinfo {author} {\bibfnamefont {Y.-S.}\ \bibnamefont {Chen}},\ }\bibfield
  {title} {\bibinfo {title} {Shell model method for gamow-teller transitions in
  heavy, deformed nuclei},\ }\href@noop {} {\bibfield  {journal} {\bibinfo
  {journal} {Phys. Rev. C}\ }\textbf {\bibinfo {volume} {74}},\ \bibinfo
  {pages} {054303} (\bibinfo {year} {2006})}\BibitemShut {NoStop}%
\bibitem [{\citenamefont {Wang}\ \emph {et~al.}(2014)\citenamefont {Wang},
  \citenamefont {Chen}, \citenamefont {Mizusaki}, \citenamefont {Oi},\ and\
  \citenamefont {Sun}}]{Wang-2014-R}%
  \BibitemOpen
  \bibfield  {author} {\bibinfo {author} {\bibfnamefont {L.-J.}\ \bibnamefont
  {Wang}}, \bibinfo {author} {\bibfnamefont {F.-Q.}\ \bibnamefont {Chen}},
  \bibinfo {author} {\bibfnamefont {T.}~\bibnamefont {Mizusaki}}, \bibinfo
  {author} {\bibfnamefont {M.}~\bibnamefont {Oi}},\ and\ \bibinfo {author}
  {\bibfnamefont {Y.}~\bibnamefont {Sun}},\ }\bibfield  {title} {\bibinfo
  {title} {Toward extremes of angular momentum: Application of the pfaffian
  algorithm in realistic calculations},\ }\href@noop {} {\bibfield  {journal}
  {\bibinfo  {journal} {Phys. Rev. C}\ }\textbf {\bibinfo {volume} {90}},\
  \bibinfo {pages} {011303(R)} (\bibinfo {year} {2014})}\BibitemShut {NoStop}%
\bibitem [{\citenamefont {Wang}\ \emph {et~al.}(2016)\citenamefont {Wang},
  \citenamefont {Sun}, \citenamefont {Mizusaki}, \citenamefont {Oi},\ and\
  \citenamefont {Ghorui}}]{Wang_2016_PRC}%
  \BibitemOpen
  \bibfield  {author} {\bibinfo {author} {\bibfnamefont {L.-J.}\ \bibnamefont
  {Wang}}, \bibinfo {author} {\bibfnamefont {Y.}~\bibnamefont {Sun}}, \bibinfo
  {author} {\bibfnamefont {T.}~\bibnamefont {Mizusaki}}, \bibinfo {author}
  {\bibfnamefont {M.}~\bibnamefont {Oi}},\ and\ \bibinfo {author}
  {\bibfnamefont {S.~K.}\ \bibnamefont {Ghorui}},\ }\bibfield  {title}
  {\bibinfo {title} {Reduction of collectivity at very high spins in
  $^{134}$\text{Nd}: Expanding the projected-shell-model basis up to
  10-quasiparticle states},\ }\href
  {https://doi.org/10.1103/PhysRevC.93.034322} {\bibfield  {journal} {\bibinfo
  {journal} {Phys. Rev. C}\ }\textbf {\bibinfo {volume} {93}},\ \bibinfo
  {pages} {034322} (\bibinfo {year} {2016})}\BibitemShut {NoStop}%
\bibitem [{\citenamefont {Sun}(2016)}]{PSM-Sun2}%
  \BibitemOpen
  \bibfield  {author} {\bibinfo {author} {\bibfnamefont {Y.}~\bibnamefont
  {Sun}},\ }\bibfield  {title} {\bibinfo {title} {Projection techniques to
  approach the nuclear many-body problem},\ }\href@noop {} {\bibfield
  {journal} {\bibinfo  {journal} {Phys. Scr.}\ }\textbf {\bibinfo {volume}
  {91}},\ \bibinfo {pages} {043005} (\bibinfo {year} {2016})}\BibitemShut
  {NoStop}%
\bibitem [{\citenamefont {Wang}\ \emph
  {et~al.}(2018{\natexlab{b}})\citenamefont {Wang}, \citenamefont {Sun},\ and\
  \citenamefont {Ghorui}}]{Wang2018}%
  \BibitemOpen
  \bibfield  {author} {\bibinfo {author} {\bibfnamefont {L.-J.}\ \bibnamefont
  {Wang}}, \bibinfo {author} {\bibfnamefont {Y.}~\bibnamefont {Sun}},\ and\
  \bibinfo {author} {\bibfnamefont {S.~K.}\ \bibnamefont {Ghorui}},\ }\bibfield
   {title} {\bibinfo {title} {Shell-model method for gamow-teller transitions
  in heavy deformed odd-mass nuclei},\ }\href@noop {} {\bibfield  {journal}
  {\bibinfo  {journal} {Phys. Rev. C}\ }\textbf {\bibinfo {volume} {97}},\
  \bibinfo {pages} {044302} (\bibinfo {year} {2018}{\natexlab{b}})}\BibitemShut
  {NoStop}%
\bibitem [{\citenamefont {Ring}\ and\ \citenamefont
  {Schuck}(1980)}]{many-body}%
  \BibitemOpen
  \bibfield  {author} {\bibinfo {author} {\bibfnamefont {P.}~\bibnamefont
  {Ring}}\ and\ \bibinfo {author} {\bibfnamefont {P.}~\bibnamefont {Schuck}},\
  }\href@noop {} {\emph {\bibinfo {title} {The nuclear many-body problem}}}\
  (\bibinfo  {publisher} {Springer-Verlag},\ \bibinfo {year}
  {1980})\BibitemShut {NoStop}%
\bibitem [{\citenamefont {Hara}\ and\ \citenamefont {Sun}(1995)}]{PSM-review}%
  \BibitemOpen
  \bibfield  {author} {\bibinfo {author} {\bibfnamefont {K.}~\bibnamefont
  {Hara}}\ and\ \bibinfo {author} {\bibfnamefont {Y.}~\bibnamefont {Sun}},\
  }\bibfield  {title} {\bibinfo {title} {Projected shell model and high-spin
  spectroscopy},\ }\href@noop {} {\bibfield  {journal} {\bibinfo  {journal}
  {Int. J. Mod. Phys. E}\ }\textbf {\bibinfo {volume} {4}},\ \bibinfo {pages}
  {637} (\bibinfo {year} {1995})}\BibitemShut {NoStop}%
\bibitem{Nilsson_RMP} A. K. Jain et al., Rev. Mod. Phys. \textbf{62}, 393 (1990).
\bibitem [{\citenamefont {Wang}\ \emph {et~al.}(2019)\citenamefont {Wang},
  \citenamefont {Dong}, \citenamefont {Chen},\ and\ \citenamefont
  {Sun}}]{Wang_2019_JpG}%
  \BibitemOpen
  \bibfield  {author} {\bibinfo {author} {\bibfnamefont {L.-J.}\ \bibnamefont
  {Wang}}, \bibinfo {author} {\bibfnamefont {J.}~\bibnamefont {Dong}}, \bibinfo
  {author} {\bibfnamefont {F.-Q.}\ \bibnamefont {Chen}},\ and\ \bibinfo
  {author} {\bibfnamefont {Y.}~\bibnamefont {Sun}},\ }\bibfield  {title}
  {\bibinfo {title} {Projected shell model analysis of structural evolution and
  chaoticity in fast-rotating nuclei},\ }\href
  {https://doi.org/10.1088/1361-6471/ab33be} {\bibfield  {journal} {\bibinfo
  {journal} {J. Phys. G}\ }\textbf {\bibinfo {volume} {46}},\ \bibinfo {pages}
  {105102} (\bibinfo {year} {2019})}\BibitemShut {NoStop}%
\bibitem [{\citenamefont {Wang}\ \emph {et~al.}(2020)\citenamefont {Wang},
  \citenamefont {Chen},\ and\ \citenamefont {Sun}}]{Wang_2020_PLB_chaos}%
  \BibitemOpen
  \bibfield  {author} {\bibinfo {author} {\bibfnamefont {L.-J.}\ \bibnamefont
  {Wang}}, \bibinfo {author} {\bibfnamefont {F.-Q.}\ \bibnamefont {Chen}},\
  and\ \bibinfo {author} {\bibfnamefont {Y.}~\bibnamefont {Sun}},\ }\bibfield
  {title} {\bibinfo {title} {Basis-dependent measures and analysis
  uncertainties in nuclear chaoticity},\ }\href
  {https://doi.org/https://doi.org/10.1016/j.physletb.2020.135676} {\bibfield
  {journal} {\bibinfo  {journal} {Phys. Lett. B}\ }\textbf {\bibinfo {volume}
  {808}},\ \bibinfo {pages} {135676} (\bibinfo {year} {2020})}\BibitemShut
  {NoStop}%
\bibitem [{\citenamefont {Guttormsen}\ \emph {et~al.}(2015)\citenamefont
  {Guttormsen}, \citenamefont {Aiche}, \citenamefont {Bello~Garrote},
  \citenamefont {Bernstein}, \citenamefont {Bleuel}, \citenamefont {Byun},
  \citenamefont {Ducasse}, \citenamefont {Eriksen}, \citenamefont {Giacoppo},
  \citenamefont {Görgen}, \citenamefont {Gunsing}, \citenamefont {Hagen},
  \citenamefont {Jurado}, \citenamefont {Klintefjord}, \citenamefont {Larsen},
  \citenamefont {Lebois}, \citenamefont {Leniau}, \citenamefont {Nyhus},
  \citenamefont {Renstrøm}, \citenamefont {Rose}, \citenamefont {Sahin},
  \citenamefont {Siem}, \citenamefont {Tornyi}, \citenamefont {Tveten},
  \citenamefont {Voinov}, \citenamefont {Wiedeking},\ and\ \citenamefont
  {Wilson}}]{Guttormsen_2015_EPJA}%
  \BibitemOpen
  \bibfield  {author} {\bibinfo {author} {\bibfnamefont {M.}~\bibnamefont
  {Guttormsen}}, \bibinfo {author} {\bibfnamefont {M.}~\bibnamefont {Aiche}},
  \bibinfo {author} {\bibfnamefont {F.~L.}\ \bibnamefont {Bello~Garrote}},
  \bibinfo {author} {\bibfnamefont {L.~A.}\ \bibnamefont {Bernstein}}, \bibinfo
  {author} {\bibfnamefont {D.~L.}\ \bibnamefont {Bleuel}}, \bibinfo {author}
  {\bibfnamefont {Y.}~\bibnamefont {Byun}}, \bibinfo {author} {\bibfnamefont
  {Q.}~\bibnamefont {Ducasse}}, \bibinfo {author} {\bibfnamefont {T.~K.}\
  \bibnamefont {Eriksen}}, \bibinfo {author} {\bibfnamefont {F.}~\bibnamefont
  {Giacoppo}}, \bibinfo {author} {\bibfnamefont {A.}~\bibnamefont {Görgen}},
  \bibinfo {author} {\bibfnamefont {F.}~\bibnamefont {Gunsing}}, \bibinfo
  {author} {\bibfnamefont {T.~W.}\ \bibnamefont {Hagen}}, \bibinfo {author}
  {\bibfnamefont {B.}~\bibnamefont {Jurado}}, \bibinfo {author} {\bibfnamefont
  {M.}~\bibnamefont {Klintefjord}}, \bibinfo {author} {\bibfnamefont {A.~C.}\
  \bibnamefont {Larsen}}, \bibinfo {author} {\bibfnamefont {L.}~\bibnamefont
  {Lebois}}, \bibinfo {author} {\bibfnamefont {B.}~\bibnamefont {Leniau}},
  \bibinfo {author} {\bibfnamefont {H.~T.}\ \bibnamefont {Nyhus}}, \bibinfo
  {author} {\bibfnamefont {T.}~\bibnamefont {Renstrøm}}, \bibinfo {author}
  {\bibfnamefont {S.~J.}\ \bibnamefont {Rose}}, \bibinfo {author}
  {\bibfnamefont {E.}~\bibnamefont {Sahin}}, \bibinfo {author} {\bibfnamefont
  {S.}~\bibnamefont {Siem}}, \bibinfo {author} {\bibfnamefont {T.~G.}\
  \bibnamefont {Tornyi}}, \bibinfo {author} {\bibfnamefont {G.~M.}\
  \bibnamefont {Tveten}}, \bibinfo {author} {\bibfnamefont {A.}~\bibnamefont
  {Voinov}}, \bibinfo {author} {\bibfnamefont {M.}~\bibnamefont {Wiedeking}},\
  and\ \bibinfo {author} {\bibfnamefont {J.}~\bibnamefont {Wilson}},\
  }\bibfield  {title} {\bibinfo {title} {Experimental level densities of atomic
  nuclei},\ }\href {https://doi.org/10.1140/epja/i2015-15170-4} {\bibfield
  {journal} {\bibinfo  {journal} {Eur. Phys. J. A.}\ }\textbf {\bibinfo
  {volume} {51}},\ \bibinfo {pages} {170} (\bibinfo {year} {2015})}\BibitemShut
  {NoStop}%
\bibitem [{\citenamefont {Baglin}(2011)}]{93_level_data}%
  \BibitemOpen
  \bibfield  {author} {\bibinfo {author} {\bibfnamefont {C.~M.}\ \bibnamefont
  {Baglin}},\ }\bibfield  {title} {\bibinfo {title} {Nuclear data sheets for
  $\mathrm{A} = 93$},\ }\href
  {https://doi.org/https://doi.org/10.1016/j.nds.2011.04.001} {\bibfield
  {journal} {\bibinfo  {journal} {Nuclear Data Sheets}\ }\textbf {\bibinfo
  {volume} {112}},\ \bibinfo {pages} {1163 } (\bibinfo {year}
  {2011})}\BibitemShut {NoStop}%
\bibitem [{\citenamefont {Misch}\ \emph {et~al.}(shed)\citenamefont {Misch},
  \citenamefont {Sprouse}, \citenamefont {Mumpower}, \citenamefont {Couture},
  \citenamefont {Fryer}, \citenamefont {Meyer},\ and\ \citenamefont
  {Sun}}]{Wendell_2021_tmp}%
  \BibitemOpen
  \bibfield  {author} {\bibinfo {author} {\bibfnamefont {G.~W.}\ \bibnamefont
  {Misch}}, \bibinfo {author} {\bibfnamefont {T.~M.}\ \bibnamefont {Sprouse}},
  \bibinfo {author} {\bibfnamefont {M.~R.}\ \bibnamefont {Mumpower}}, \bibinfo
  {author} {\bibfnamefont {A.}~\bibnamefont {Couture}}, \bibinfo {author}
  {\bibfnamefont {C.~L.}\ \bibnamefont {Fryer}}, \bibinfo {author}
  {\bibfnamefont {B.~S.}\ \bibnamefont {Meyer}},\ and\ \bibinfo {author}
  {\bibfnamefont {Y.}~\bibnamefont {Sun}},\ }\href@noop {} {\bibfield
  {journal} {\bibinfo  {journal} {Symmetry}\ } \textbf {\bibinfo
  {volume} {13}},\ \bibinfo {pages} {1831 } (\bibinfo {year}
  {2021})}\BibitemShut {NoStop}%
\end{thebibliography}

%

\end{document}